\documentclass[sigconf]{acmart}

\pdfoutput=1

\usepackage{booktabs}

\usepackage{framed}
\usepackage{hyperref}
\usepackage{gensymb}
\usepackage{subcaption}
\usepackage{tabularx}
\newcolumntype{Y}{>{\raggedright\arraybackslash}X}
\graphicspath{ {imgs/} }

\usepackage[ruled]{algorithm2e} 

\SetAlFnt{\small}
\SetAlCapFnt{\small}
\SetAlCapNameFnt{\small}
\SetAlCapHSkip{0pt}
\IncMargin{-\parindent}

\usepackage{pdfpages}
\usepackage[T1]{fontenc} 

\def\plaintitle{Privacy-Aware Eye Tracking Using Differential Privacy}

\def\plainkeywords{Online Survey; Data Sharing; Privacy Protection; Gaze Behaviour; Eye Movements; User Modeling}

 \title[Privacy-Aware Eye Tracking Using Differential Privacy]{\plaintitle}

\definecolor{linkColor}{RGB}{6,125,233}

\usepackage{amsthm} 
\newtheorem{mydef}{Definition}
\newtheorem{mythrm}{Theorem}

\renewcommand{\textrightarrow}{$\rightarrow$}

\usepackage{changes}
\colorlet{Changes@Color}{red}

\begin{document}

\copyrightyear{2019}
\acmYear{2019}
\setcopyright{acmlicensed}
\acmConference[ETRA '19]{2019 Symposium on Eye Tracking Research and Applications}{June 25--28, 2019}{Denver, CO, USA}
\acmBooktitle{2019 Symposium on Eye Tracking Research and Applications (ETRA '19), June 25--28, 2019, Denver, CO, USA}
\acmPrice{15.00}
\acmDOI{10.1145/3314111.3319915}
\acmISBN{978-1-4503-6709-7/19/06}

\sloppy

\title{\plaintitle}

\author{Julian Steil}
\affiliation{
\institution{Max Planck Institute for Informatics\\ Saarland Informatics Campus}
 }
\email{jsteil@mpi-inf.mpg.de}

\author{Inken Hagestedt}
\affiliation{\institution{CISPA Helmholtz Center for Information Security\\ Saarland Informatics Campus}
 }
\email{inken.hagestedt@uni-saarland.de}

\author{Michael Xuelin Huang}
\affiliation{\institution{Max Planck Institute for Informatics\\ Saarland Informatics Campus}
 }
\email{mhuang@mpi-inf.mpg.de}

\author{Andreas Bulling}
\affiliation{\institution{
University of Stuttgart\\ Institute for Visualisation and Interactive Systems}
 }
\email{andreas.bulling@vis.uni-stuttgart.de}

\begin{abstract}
With eye tracking being increasingly integrated into virtual and augmented reality (VR/AR) head-mounted displays, preserving users' privacy is an ever more important, yet under-explored, topic in the eye tracking community.
We report a large-scale online survey (N=124) on privacy aspects of eye tracking that provides the first comprehensive account of with whom, for which services, and to what extent users are willing to share their gaze data.
Using these insights, we design a privacy-aware VR interface that uses differential privacy, which we evaluate on a new 20-participant dataset for two privacy sensitive tasks: 
We show that our method can prevent user re-identification and protect gender information while maintaining high performance for gaze-based document type classification.
Our results highlight the privacy challenges particular to gaze data and demonstrate that differential privacy is a potential means to address them. Thus, this paper lays important foundations for future research on privacy-aware gaze interfaces.
\vspace{-0.2cm}
\end{abstract}



\keywords{\plainkeywords \vspace{-0.05cm}}

\maketitle

\newcommand{\mechanism}{\mathcal{M}}
\newcommand{\database}{D}
\newcommand{\aggregate}{g} %f is reserved for ``feature'' ;)
\newcommand{\outdim}{d} %output dimension of g
\newcommand{\sensitivity}[1]{\Delta_{#1}}
\newcommand{\noise}{\eta} 
\newcommand{\f}{f} %feature - I admit, it looks stupid, but this way it's clear that the letter 'f' is assigned to the feature and can not be used in a different context
\newcommand{\p}[1]{p_{#1}} %participant (with various indices)
\newcommand{\ti}{t} %time index - \t is already defined and I want to minimize typing
\newcommand{\timax}{tmax}
\newcommand{\range}[1]{\delta_{#1}} % range of feature
\newcommand{\utility}{u} %utility for exponential mechanism
\newcommand{\sampledOutput}{r} 
\newcommand{\sampledFrom}{\sim} %I just can't remember the latex code for this symbol
\newcommand{\y}{y} %the one sample we get from exponential distribution
\newcommand{\subsampleWindow}{w}
\newcommand{\numFeatures}{m} %number of participants in dataset is n, by the way
\vspace{-0.1cm}
\section{Introduction}

\begin{figure}[t!]
	\vspace{0.2cm}
    \centering
    \includegraphics[width=1\columnwidth]{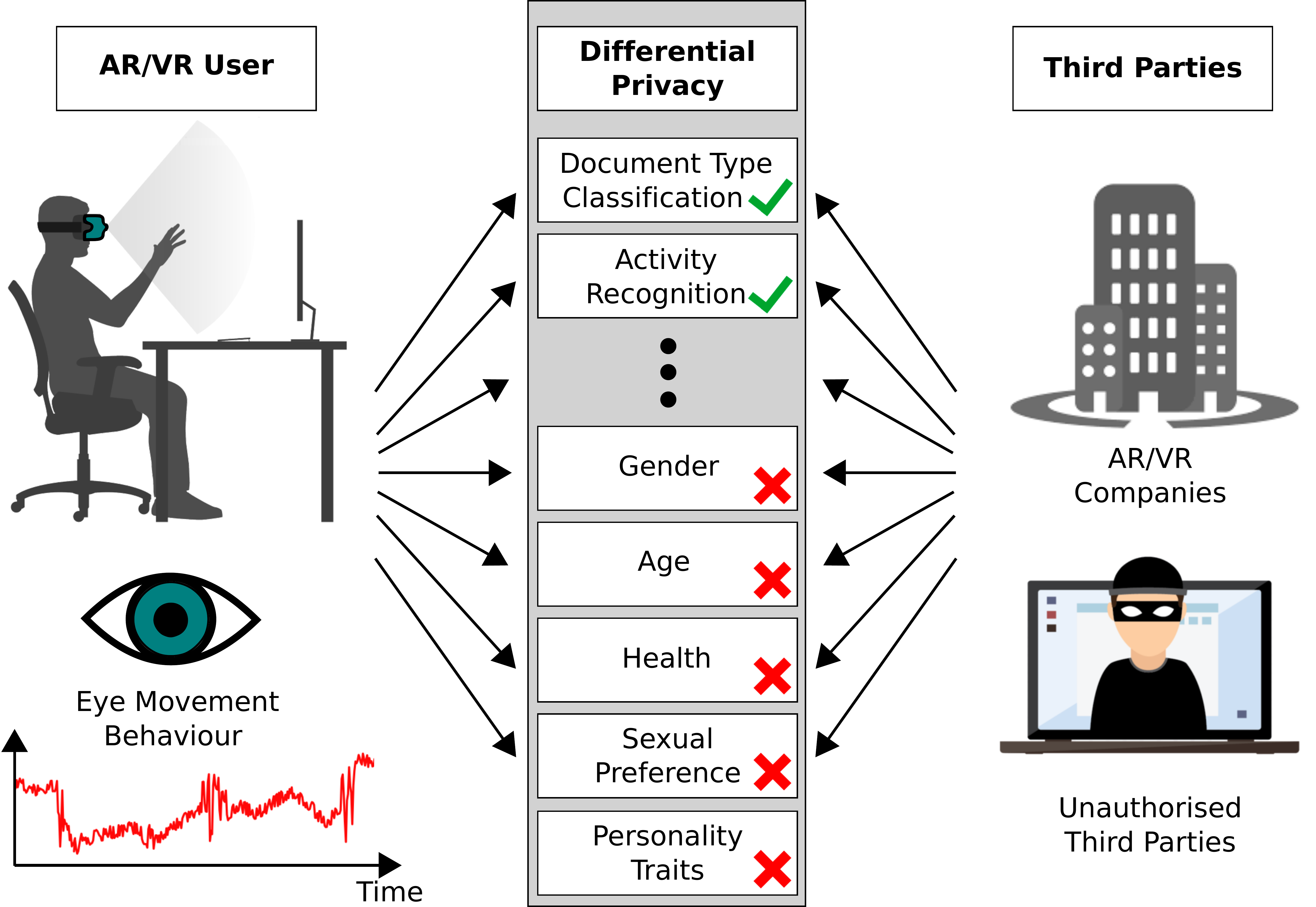}
    \vspace{-0.5cm}
    \caption{Using differential privacy prevents third parties, like companies or hackers, from deriving private attributes from a user's eye movement behaviour while maintaining the data utility for non-private information.}
    \label{fig:teaser}
\end{figure}

With eye tracking becoming pervasive~\cite{bulling10_pcm,tonsen2017invisibleeye}, preserving users' privacy has emerged as an important topic in the eye tracking, eye movement analysis, and gaze interaction research communities.
Privacy is particularly important in this context given the rich information content available in human eye movements~\cite{bulling11_pcm}, on one hand, and the rapidly increasing capabilities of interactive systems to sense, analyse, and exploit this information in \mbox{everyday} life~\cite{hansen2003command,stellmach2012look,vertegaal2003attentive} on the other.
The eyes are more privacy-sensitive than other input modalities: They are typically not consciously controlled; they can reveal unique private information, such as personal preferences, goals, or intentions. 
Moreover, eye movements are difficult to remember, let alone reconstruct in detail, in retrospect, and hence do not easily allow users to ``learn from their mistakes'', i.e.\ to reflect on their past and change their future privacy-related behaviour.

These unique properties and rapid technological advances call for new research on next-generation eye tracking systems that are \textit{privacy-aware}, i.e.\ that preserve users' privacy in all \mbox{interactions} they perform with other humans or computing \mbox{systems} in \mbox{everyday} life.
However, \textit{privacy-aware eye tracking} remains under-investigated as of yet~\cite{liebling2014privacy}.

The lack of research on privacy-aware eye tracking results in two major limitations:
First, there is a lack of even basic understanding of users' privacy concerns with eye tracking in general and eye movement analysis in particular.
Second, there is a lack of eye tracking methods to preserve users' privacy, corresponding systems, and user interfaces that implement (and hence permit the \mbox{evaluation} of) these methods with end users.
Our work aims to address both limitations and, as such, make the first crucial step towards a new generation of eye tracking systems that respect and actively \mbox{protect private information that can be inferred from the eyes.}

\textit{Our work first contributes a large-scale online survey on \mbox{privacy} aspects of eye tracking and eye movement analysis}. 
The survey provides the first comprehensive account of with whom, for which services, and to what extent users are willing to share their eye movement data.
The survey data is available at \url{https://www.mpi-inf.mpg.de/MPIIDPEye/}.
Informed by the survey, \textit{we further contribute the first method to protect users' privacy in eye tracking based on differential privacy (DP)}, a well-studied framework in the privacy research community.
In a nutshell, DP adds noise to the data so as to minimise chances to infer privacy-sensitive information or to (re-)identify a user while, at the same time, still allow use of the data for desired applications (the so-called utility task), such as activity recognition or document type classification (see Figure~\ref{fig:teaser}).
We illustrate the use of differential privacy for a sample virtual reality (VR) gaze interface.
We opted for a VR interface given that eye tracking will be readily integrated into upcoming VR head-mounted displays,
and hence, given the significant and imminent threat potential~\cite{adams2018ethics}:
Eye movement data may soon be collected at scale on these devices, recorded in the background without the user noticing, or even transferred to hardware manufacturers.
\section{Related Work}

We discuss previous works on 1) information available in eye movements, 2) eye movements as a biometric, and 3) differential privacy.

\subsection{Information Available in Eye Movements}

A large body of work across different research fields has demonstrated the rich information content available in human eye movements.
Pupil size is related to a person's interest in a scene~\cite{hess1960pupil} and can be used to measure cognitive load~\cite{matthews1991pupillary}.
Other works have shown that eye movements are closely linked to mental disorders, such as Alzheimer's 
~\cite{hutton1984eye}, Parkinson's 
~\cite{kuechenmeister1977eye}, or schizophrenia~\cite{holzman1974eye}.
More recent work in HCI has demonstrated the use of eye movement analysis for human activity recognition~\cite{bulling13_chi,steil2015discovery} as well as to infer a user's cognitive state~\cite{bulling14_pcm,faber2017automated} or personality traits~\cite{hoppe2018eye}.
More closely related to our work, several researchers have shown that gender and age can be inferred from eye movements, e.g.\ by analysing the spatial distribution of gaze on images like faces~\cite{sammaknejad2017gender,cantoni2015gant}.

All of these works underline the significant potential of eye movement analysis for a range of future applications,
some of which may soon become a reality,
for example, with the advent of eye tracking-equipped virtual and augmented reality head-mounted displays.
Despite the benefits of these future applications, the wide availability of eye tracking will also pose significant privacy risks that remain under-explored in the eye tracking community.

\subsection{Eye Movements as a Biometric}

Eye movement biometrics has emerged as a promising approach to user authentication~\cite{kasprowski2003eye}.
While first works required a point stimulus that users were instructed to follow with their eyes~\cite{kasprowski2004human,kasprowski2005enhancing}, later ones explored static points~\cite{bednarik2005eye} or images~\cite{maeder2003visual}.
Kinnunen et al.\ presented the first method for ``task-independent'' person authentication using eye movements~\cite{kinnunen2010towards}.
Komogortsev et al.\ proposed the first attempt to model eye movements for authentication using an Oculomotor Plant Mathematical Model~\cite{komogortsev2010biometric,komogortsev2013biometric}.
Eberz et al.\ presented a biometric based on eye movement patterns.
They used 20 features that allowed them to reliably distinguish and authenticate users across a variety of real-world tasks, including reading, writing, web browsing, and watching videos on a desktop screen~\cite{eberz2016looks}.
\mbox{Zhang et al.} used eye movements to continuously authenticate the wearer of a VR headset by showing different visual stimuli~\cite{zhang2018continuous}.

While an ever-growing body of research explores eye movements as a promising modality for privacy applications and user authentication, we are the first to practically explore eye \mbox{movements} recorded using eye tracking as a potential threat to users' privacy.

\subsection{Differential Privacy}

Differential privacy has been studied in privacy research for more than a decade
in terms of its theoretical foundations and its practical applications to different data types, such as 
location~\cite{pyrgelis2017knock},
biomedical data~\cite{saleheen2016msieve}, or continuous time series data~\cite{fan2012adaptively}. 
We refer the reader to~\cite{zhu2017differentially} for a survey.
A key challenge in differential privacy is to find the right trade-off between privacy and utility, that is, the right amount of random noise to ``hide'' an individual without hampering data utility.
Fredrikson et al.\ demonstrated how important it is to balance privacy and utility~\cite{fredrikson2014privacy}.
They observed that either privacy was not preserved or that utility suffered, leading to increased health risks for the patients from unsuitable drug dosage. 
A good privacy-utility trade-off is possible if privacy mechanisms are tailored towards a specific use case~\cite{pyrgelis2017knock,fan2012adaptively}.
While differential privacy has a long history in privacy research, to the best of our knowledge, we are the first to apply this framework to eye tracking data.
\section{Privacy Concerns in Eye Tracking}

\begin{figure*}[t!]
  \vspace{-0.4cm}
    \centering
        \centering
        \includegraphics[width=0.92\textwidth]{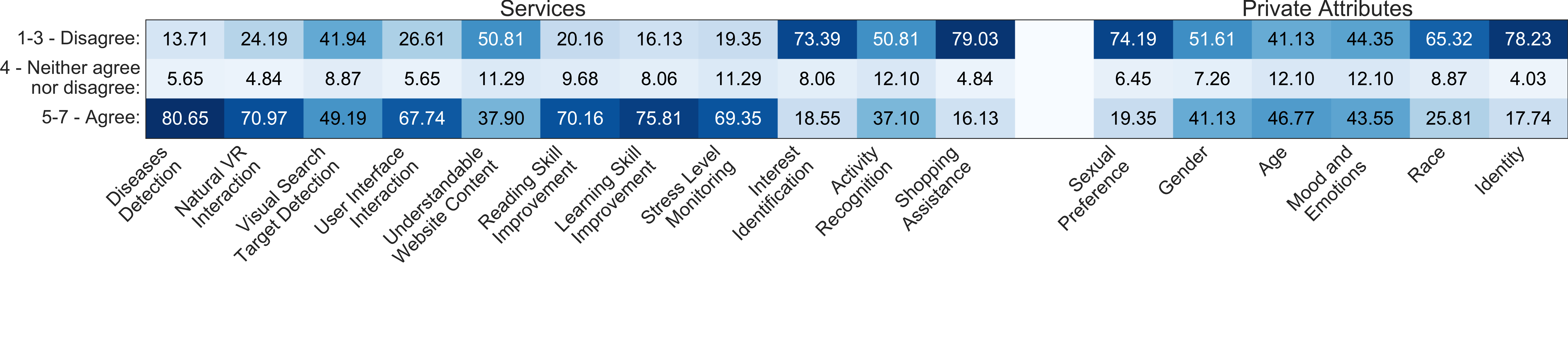} 
        \vspace{-4.0em}
         \caption{Survey results (Services and Attributes): With which services would you agree to share your eye tracking data (Services)?; Would you agree to private attributes being inferred by these services (Private Attributes)?  
        }
        \label{fig:services}
\end{figure*}

\begin{figure*}[t!]
\vspace{-0.35cm}
    \centering
        \centering
        \includegraphics[width=0.92\textwidth]{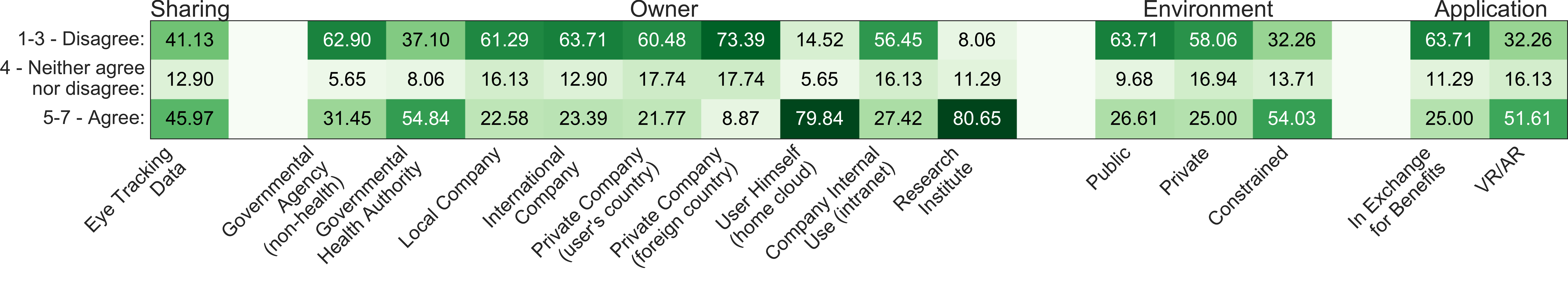} 
        \vspace{-1.8em}
        \caption{Survey results (Whom and Where): Would you agree to share your eye tracking data in general (Sharing); with whom (Owner); where (Environment); in exchange for benefits or for VR/AR usage (Application)?
        }
        \label{fig:owner}
        \vspace{-0.2cm}
\end{figure*}

We conducted a large-scale online survey to shed light on users' privacy concerns related to eye tracking technology 
and the information that can be inferred from eye movement data.
We advertised our survey on social platforms (Facebook, WeChat) and local mailing lists for study announcements.
The survey opened with general questions about eye tracking and VR technologies; continued with questions about future use and applications, data sharing and \mbox{privacy} (especially regarding with whom users are willing to share their data); and concluded with questions about the participants' willingness to share different eye movement representations. 
Participants answered each question on a 7-point Likert scale (1: Strongly disagree to 7: Strongly agree). 
To simplify the analysis, we merged scores 1 to 3 to ``Disagree'' and 5 to 7 to ``Agree''.

The survey took about 20 minutes to complete, was set up as a Google Form, and was split into the parts described above.
Our design ensured that 
participants without pre-knowledge of eye tracking and VR technology could participate as well:
We provided a slide show containing 
information about eye tracking in general, and in VR devices specifically, 
and
introduced the different forms of data representation, showing example images or explanatory texts. %an extended explanation.

\begin{figure*}[t!]
 \vspace{-0.4cm}
    \centering
    \includegraphics[width=1.0\textwidth]{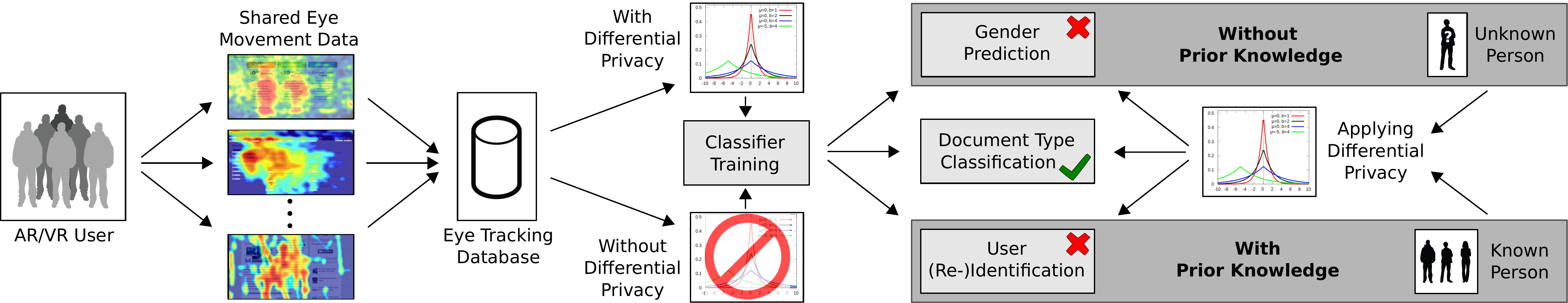}
    \vspace{-0.7cm}
     \caption{
     (Left) Our method assumes that AR/VR users share their eye tracking data and privacy-sensitive information with a third party, which is able to train classifiers with or without differentially private data to infer private attributes of an \mbox{unknown} (without prior knowledge) or a known (with prior knowledge) person; (Right) Applying differential privacy to \mbox{test data} prevents private information inference (gender, user (re-)identification) but maintains data utility (document type classification).
     }
    \label{fig:method}
\vspace{-0.1cm}
\end{figure*}

In our survey, 124 people (81 male, 39 female, 4 did not tick the gender box) participated, aged 21 to 66 (mean = 28.07, std = 5.89).
The participants were from all over the world, coming from 29 different countries (Germany: 39\%, India: 12\%, \mbox{Pakistan:} 6\%, Italy: 6\%, China: 5\%, USA: 3\%). Sixty-seven percent of them had a graduate university degree (master's or PhD), and 22\% had an undergraduate university degree (bachelor's).
Fifty-one percent were students of a variety of subjects (law, language science, computer science, psychology, etc.); 34\% were scientists and researchers, IT professionals (7\%), or had business administration jobs (2\%).
Since the topic of the survey was in the title of posts and emails, most likely people inherently interested in the topic participated. The majority were young, educated people with a technical background the exact group of people most likely to experience AR/VR technology (73\%) in contrast to, for example, older generations.

Given the breadth of results, we highlight key insights most relevant for the current paper. 
We found nearly all answers for the provided questions to be significantly different from an equal distribution tested with Pearson's chi-squared test \mbox{(p < 0.001, dof = 6)}.
Additionally, we calculated the skewness and observed that the majority of questions show a significant difference to the corresponding normal distribution (p < 0.1).
Detailed numbers, plots, significance and skewness test results can be found in the supplementary material (see \url{https://www.mpi-inf.mpg.de/MPIIDPEye/}).

\paragraph{Services and Attributes:}
In the first part of our survey, we asked participants for which services they would share their eye tracking data 
and presented both currently 
available and potential future services as answer options.
As we can see from Figure~\ref{fig:services}, more than 80\% of all participants agreed to share their eye tracking data for (early) detection of diseases like Alzheimer's or Parkinson's. Likewise, the majority agreed to share their data for hands-free VR and user interface interaction. 
Similar results can be observed for learning and reading skill detection as well as for stress level monitoring. 
However, for improved gaze target recognition, website content, and activity recognition, we observe two peaks.
A clear majority is unwilling to share data with shopping assistance and interest detection services. 

Our next set of questions indicated the fact that services could be able to infer private attributes from their data, and we asked whether
participants would still want to share their eye tracking data.
We clearly observed that if the attributes of sexual preference, gender, race, and identity can be inferred, 
a majority
do not want to share their data. 
It was only for age and emotion detection that we identified two different interest groups that either agree with or object to sharing their data.

\paragraph{Whom and Where:}
In the second part, of our survey we asked participants whether they would share eye tracking data in general, and with whom.
Moreover, we were interested in whether the environment has an influence on their sharing behaviour (see Figure~\ref{fig:owner}). 
Finally, we wanted to know whether the sharing behaviour is different if participants get benefits (not 
specified) in exchange for their data 
or if the data is collected during VR/AR usage in general.

The answers as to whether participants would share their eye tracking data in general do not show a clear tendency; the participants' opinions are split in two groups ($\chi^{2}$(dof = 6) = 32.25, p = $1.46\times10^{-6}$).
Next, we asked more specifically whether participants would share their data if it were later owned and operated by one of the given ``owner'' options in Figure~\ref{fig:owner}. 
According to their answers, participants would only share their data if the co-owner is a governmental health-agency; they do not trust local and international companies, or company internal use. 
However, participants would also  
share their data for research purposes, which is not surprising given that 67\% of participants have a graduate university degree and trust in research institutes.
Participants would not agree to share their data in public, nor in private environments, but they would agree to constrained environments. 
Furthermore, the participants object to sharing their data for any kind of benefit, but would agree 
when their eye tracking data was collected in 
VR/AR \mbox{($\chi^{2}$(dof = 6) = 26.72, p = 0.00016)}.

\paragraph{Data Representation:}
In the final part of the survey, we asked participants in what form they would agree to share their data. We discriminate 12 different representations, ranging from raw eye tracking, to heatmaps, to aggregated features (see Figure 3 in the supplementary material).
Additionally, we were interested in whether their sharing behaviour changes if the data is first anonymised. 
Information which provides gaze information, like fixations, or scan path information on a surface would mostly not be shared.
Participants largely agree to share their eye tracking data as statistical features, and especially aggregated features. 
This is why we focus in our study on the aggregated feature representation to apply differential privacy.
Our survey shows a clear increase in participants willing to share their data in anonymised form.
\section{Privacy-preserving Eye Tracking}

The findings from our survey underline the urgent need to develop 
\textit{privacy-aware eye tracking systems} -- systems that 
provide a formal guarantee to protect the privacy of their users. 
Additionally, it is important not to forget that eye movement data typically also serves a desired task -- a so-called \textit{utility}.
For example, eye movement data may be used in a reading assistant to detect 
the documents a user is reading~\cite{kunze2013know} or to automatically estimate how many words a user reads per day~\cite{kunze2013wordometer,kunze2015quantifying}. 
Therefore, it is important to ensure that any privacy-preserving method does not render the utility dysfunctional, i.e.\ that the performance on the utility task will not drop too far.
The key challenge can thus be described as \textit{ensuring privacy without impeding utility}.

We assume in the following that multiple users share their eye tracking data in the form of aggregated features.
The resulting eye tracking database is visualised in the left part of Figure~\ref{fig:method}.
This database can be downloaded both for legitimate use cases as well as for infringing on users' privacy, for example, to train classifiers for various tasks.
Therefore, our proposed privacy mechanism is applied prior to the release by a trusted curator. 

\subsection{Threat Models}

We have identified two attack vectors on users' privacy in the context of eye tracking that we formalise in two threat models.
They differ in their assumption about the attackers' prior knowledge about their target
(see the right part of Figure~\ref{fig:method}).
\paragraph{Without Prior Knowledge}
In the first threat model, we assume that an attacker has no prior knowledge about the target and wants to infer a private attribute; 
we focus on gender in our example study.
The attacker can only rely on a training data set from multiple participants different from the target.
This data can be gathered by companies or game developers we share our data with in exchange for a specific service.
Some users might opt in to share their data with a third party to receive personalised advertisements, or they might create a user account to remove advertisements.
These companies with eye tracking data can misuse the data, forward it to third parties or get hacked by external attackers.
Another source for attackers to get eye tracking datasets is publicly available datasets generated for research purposes.
Concretely, VR glasses are offered in gaming centres and used by multiple visitors, which we refer to as the one-device-multiple-users scenario.
An attacker with access to the eye tracking data might be interested in inferring the gender of the current user
to show gender-specific advertisements. 

\paragraph{With Prior Knowledge}
The second threat model assumes that the attacker has already gathered prior knowledge about the target.
Observing further eye tracking data, the attacker wants to re-identify the target 
to inspect the target's habits.
Concretely, the target might be using different user accounts or even different devices for work and leisure time 
(a one-user-multiple-devices scenario).
We assume the attacker is able to link the target's work data to the target's identity and now 
wants to identify the target's data
from his/her leisure activities.
Again, the attacker could be a VR/AR company exploiting their data to check whether a device is only used by one person, or re-identify a user automatically to adapt device settings. Moreover, data could be released intentionally to a third party for money or unintentionally through a hack.

\subsection{Differential Privacy for Eye Tracking}

We propose to mitigate the privacy threats emerging from our two threat models using \textit{differential privacy}, a well-known framework from privacy research~\cite{dwork2014algorithmic}. 
Differential privacy guarantees that the answer of the privacy-preserving mechanism does not depend on
whether a single user contributed his/her data or not; hence, there is no way to infer further information about 
this user.
Concretely, the answer to the question ``What is the average fixation rate when reading a text?'' should be almost the same, whether or not a specific user, say, Alice, has contributed her data to our database of fixation rates. 
We denote a differentially private mechanism by $\mechanism$ and refer to Alice's data as a single data element 
in the database $\database$. 
Typically, $\mechanism$ adds random noise to ``hide'' each data element, which we 
will formalise in the following. 
\begin{mydef}[$\epsilon$-Differential Privacy \cite{dwork2006calibrating}] \label{def:differentialPrivacy} 
A mechanism $\mechanism$ provides $\epsilon$-differential privacy if 
for all databases $\database$, $\database'$ that differ in at most one element 
and for every $S \subseteq \text{Range}(\mechanism)$, we have 
\vspace{-0.2cm}
\begin{equation}
Pr[\mechanism(\database) \in S] \leq e^{\epsilon} \cdot Pr[\mechanism(\database') \in S].
\end{equation}
\vspace{-0.5cm}
\end{mydef}

Differential privacy allows computing an arbitrary function $\aggregate$ over the 
database, i.e. $\aggregate : \mathcal{R}^{*} \mapsto \mathcal{R}^{\outdim}$, where 
$\outdim$ denotes the dimensionality of the output of $\aggregate$. 
For our running example, $\aggregate$ would compute the average and output one number, hence $\outdim=1$.
Similarly, we could define $\aggregate$ to average over 30-second windows of fixation 
data and then output a vector of length $\outdim$. 

How much noise we have to add depends on the variance of the data between two 
arbitrary elements. Formally: 
\begin{mydef}[$L_1$ Sensitivity \cite{dwork2006calibrating}]\label{def:sensitivity}  
For all functions $\aggregate : \mathcal{R}^{*} \mapsto \mathcal{R}^{\outdim}$, 
the $L_1$ sensitivity is
the smallest number $\sensitivity{\aggregate}$ s.th. for all databases 
$\database, \database'$ differing in one element, we have
\begin{equation}
|| \aggregate(\database) - \aggregate(\database')||_{L_1} \leq \sensitivity{\aggregate}.
\end{equation}
\vspace{-0.5cm}
\end{mydef}
\noindent{Intuitively,} the sensitivity captures the maximal influence Alice's data could have on the 
answer to our query. In the worst case, for her privacy, Alice's data is an outlier, e.g. Alice is a 
very slow reader compared to all other participants. Even in this case, the difference between Alice's data
and any other entry in the database must be smaller than or equal to the sensitivity. 
The noise to ``hide'' Alice's contribution is scaled to this worst case,
ensuring Alice's privacy.

Next, we formalise the exponential mechanism that is one way to generate differentially private data:
\begin{mydef}[Exponential Mechanism \cite{dwork2014algorithmic}] \label{def:exponentialMechanism} 
The exponential mechanism selects and outputs an element $\sampledOutput \in \mathcal{R}$ in the range of permissible output elements 
with probability equal to (written: $\sampledOutput \sampledFrom$) 
\vspace{-0.3cm}
\begin{equation}
\sampledOutput \sampledFrom exp(\frac{\epsilon \cdot \utility(x, \sampledOutput) }{ 2 \sensitivity{\utility} } )
\end{equation}
where $\utility$ is a utility function judging the quality of $\sampledOutput$ with respect to the original data element $x$.
\end{mydef}	

\noindent{In} order to apply the exponential mechanism to our example database of fixation durations,
we would first need to define a utility function $\utility$ and the 
set of permissible outputs.
Valid answers to the query ``What are the average fixation rates when reading a text, 
sampled at 30 second windows?'' are vectors of length $\outdim$ containing real-numbered entries;
thus, $\mathcal{R} = \mathbb{R}_{\geq 0}^{\outdim}$.
The utility function $\utility$ is a measure of quality for the output $\sampledOutput$ with respect to the original data entry $x$.
The exponential mechanism ensures that high-quality outputs $\sampledOutput$ are generated exponentially more often than \mbox{low-quality $\sampledOutput$.}

Finally, we state one theorem that allows combining several 
differentially private mechanisms into one. 

\begin{mythrm}[Composition Theorem \cite{dwork2006calibrating}]\label{def:compositionTheorem}  %\footnote{Lemma 2.5)}
Let $\mechanism_1, \mbox{\small{...}} , \mechanism_k$ be a fixed sequence of mechanisms, where each 
mechanism $\mechanism_i$ is $\epsilon_i$-differentially private. 
Then, their joint output $\mechanism(\database) = (\mechanism_1(\database), \mbox{\small{...}} , \mechanism_k(\database))$ is $\epsilon$-differentially private for 
$\epsilon = \sum_{i=1}^{k} \epsilon_i$.
\end{mythrm}

\subsection{Implementing Differential Privacy}
\label{sec:ImplementingDP}

Our dataset contains data from $n$ participants,  
which we refer to as $\p{1}, \mbox{\small{...}}, \p{n}$. 
For each participant, we measure $\numFeatures$ features, $\f_1, \mbox{\small{...}} , \f_{\numFeatures}$ 
at different points in time.
In summary,
$\p{1, \f_7, \ti_5}$ denotes the value of the 7th feature at time 
point 5 of participant 1, and the vector $(\p{1, \f_7, \ti_0}, \mbox{\small{...}} , \p{1, \f_7, \ti_{max, 1}})$ 
contains all measurements of feature 7 for participant 1. 
Notice that the data entries available may have different lengths, i.e. $\ti_{max, 1}$, the last time 
point of participant 1, may be different from another participant's last time point, e.g. 
$\ti_{max, 2}$.

The sensitivity for our mechanism then depends on the range of the features, 
which is different across our $\numFeatures$ features. 
For example, feature $\f_{15}$ 
is the fixation duration in our dataset, and it
has an estimated range of [0.11, 2.75] seconds, while $f_{22}$, which describes the pupil diameter size, has an estimated range of [21.9, 133.9] pixels.
Therefore, we derive one privacy mechanism $\mechanism_{\f_i}$ for each feature separately and use the 
composition theorem (Theorem~\ref{def:compositionTheorem}) to combine the $\numFeatures$ mechanisms into our final mechanism.
The exponential mechanism requires a utility function $\utility$. 
We choose the $L_1$ distance for simplicity of the derivation: 
\vspace{-0.2cm}
\begin{equation} \label{eq:ourUtility}
\utility(\p{\f_i}, \sampledOutput) = \sum_{j=1}^{\ti_{max, \p{}}} |\p{\f_i, j} - \sampledOutput_{j} |
\vspace{-0.1cm}
\end{equation}
According to Definition~\ref{def:sensitivity}, the sensitivity $\sensitivity{\utility, \f_i}$ is
\begin{equation}
\sensitivity{\utility, \f_i}=\underset{\p{\f_i}, q_{\f_i}}{max} 
|| (p_{\f_i, \ti_0}, \mbox{\small{...}} , p_{\f_i, \ti_{max, p}})-(q_{\f_i, \ti_0}, \mbox{\small{...}} , q_{\f_i, \ti_{max, q}}) ||_{L_1}
\end{equation}
i.e. the maximal difference between the data vectors of two arbitrary participants 
$p$ and $q$ for the $i$-th feature.
Next, we unify the length by padding the data vector with the shorter length. 
Let $\timax$ be the maximal length: $\timax = max(\ti_{max, p}, \ti_{max, q})$.
Using this and the definition of the $L_1$ norm: 
\vspace{-0.1cm}
\begin{equation}
\sensitivity{\utility, \f_i} \leq \underset{p_{\f_i}, q_{\f_i}}{max} \sum_{j=1}^{\timax} 
| p_{\f_i, \ti_j} - q_{\f_i, \ti_j} |
= \timax \cdot \range{i}
\end{equation}
In the last step, we used the fact that we can derive the range $\range{i}$ 
of feature $\f_i$, either estimated from the data or 
by theoretic constraints.

We rely on the exponential mechanism (see Definition~\ref{def:exponentialMechanism})
to obtain a vector $\sampledOutput$ that is differentially private for each 
participant $\p{}$ and feature $\f_i$:
\vspace{-0.35cm}
\begin{equation}
\hspace{1cm}\sampledOutput \sampledFrom exp(\frac{\epsilon_i \utility(\p{\f_i}, \sampledOutput) }{ 2 \sensitivity{\utility, \f_i} } )
\overset{\text{Eq.~\ref{eq:ourUtility}}}{=} exp(\frac{\epsilon_i \sum_{j=1}^{\ti_{max, \p{}}} |\p{\f_i, j} - \sampledOutput_{j} | } {2 \cdot \timax \cdot \range{i}} )
\end{equation}
To increase readability, we define $\lambda_i = \frac{\epsilon_i}{ 2 \cdot \timax \cdot \range{i} }$,
which is constant once $i$ and $\epsilon_i$ are fixed. 
We generate such a vector $\sampledOutput$ from the exponential distribution by
first sampling
a random scalar $\y$ from the exponential distribution with location 0 and scale parameter $\frac{1}{\lambda_i}$.
We derive our differentially private vector $\sampledOutput$ from $\y$ as follows:
\vspace{-0.1cm}
\begin{equation}
\y  = exp( \lambda_i \cdot \sum_{j=1}^{\ti_{max, \p{}}} | \p{\f_i, j} - \sampledOutput_{j} | ) \Leftrightarrow
\frac{log_{e}(\y)}{\lambda_i} = \sum_{j=1}^{\ti_{max, \p{}}} | \p{\f_i, j} - \sampledOutput_{j} |
\vspace{-0.1cm}
\end{equation} 
Selecting $\sampledOutput_j = \pm \frac{log_{e}(\y)}{\lambda_i \times \timax } + \p{\f_i, j}$ fulfils 
the above constraint with randomly sampled sign. 

The privacy guarantee of the combined mechanism $\mechanism$ is, by the composition 
theorem (Theorem~\ref{def:compositionTheorem}), $\sum_{i=1}^{\numFeatures} \epsilon_i$. 

\paragraph{Subsampling}
In order to achieve a higher privacy guarantee, we propose to subsample the data. 
Given a window size $\subsampleWindow$, we draw one sample from 
$(\p{k, i, n \cdot \subsampleWindow}, \mbox{\small{...}} , \p{k, i, (n+1) \cdot \subsampleWindow})$ 
for each participant $k$ and feature $i$ independently 
where $n \in \mathbb{N}$, such that the sampling windows are non-overlapping. 
Notice that this subsampling approach and the corresponding window size are independent 
of the feature generation process.
This method decreases the sensitivity further by a factor of $\subsampleWindow$: 
$\sensitivity{u, \f_i, \subsampleWindow} \leq \frac{\timax}{\subsampleWindow} \cdot \range{i}$.
\section{Data Collection}

\begin{figure}
   \centering
    \begin{subfigure}{0.32\columnwidth}
    \centering
        \includegraphics[width=0.98\textwidth]{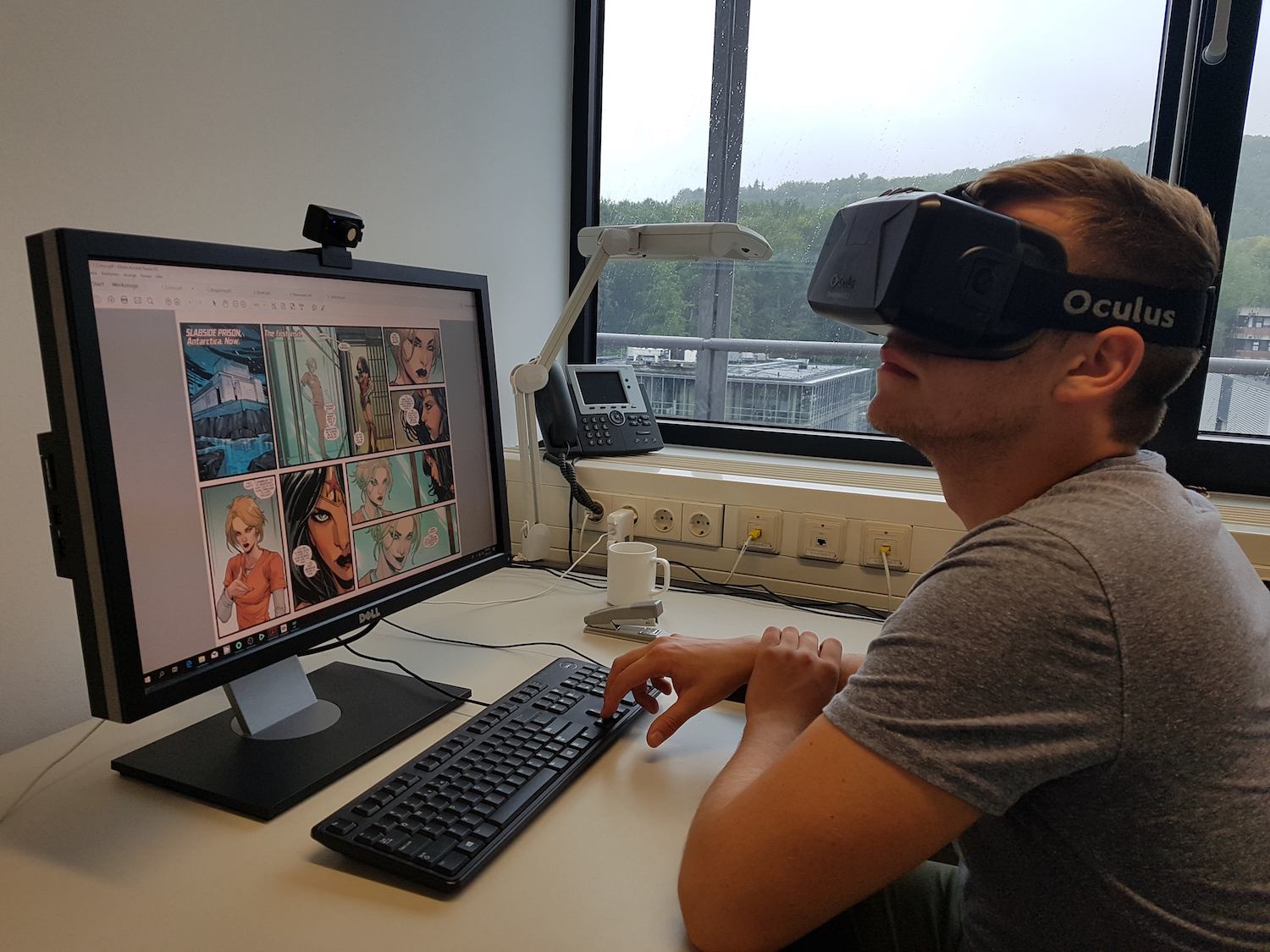}
        \vspace{-0.4cm}
        \caption{Comic}
        \label{fig:Comic}
    \end{subfigure}
    \hfill
    \begin{subfigure}{0.32\columnwidth}
    \centering
\includegraphics[width=0.98\textwidth]{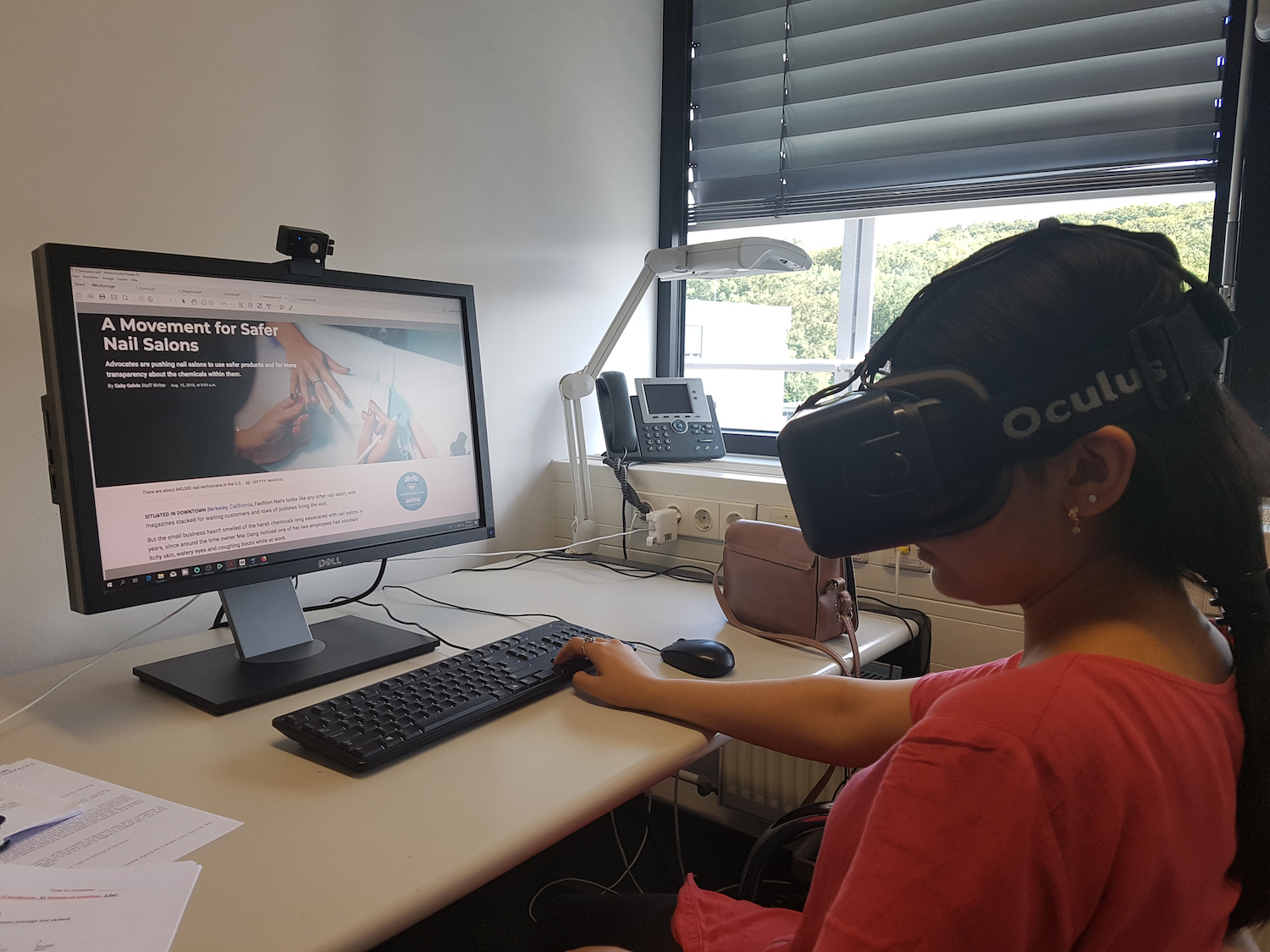}
        \vspace{-0.4cm}
        \caption{Newspaper}
        \label{fig:Newspaper}
    \end{subfigure}
    \hfill
    \begin{subfigure}{0.32\columnwidth}
    \centering
        \includegraphics[width=0.98\textwidth]{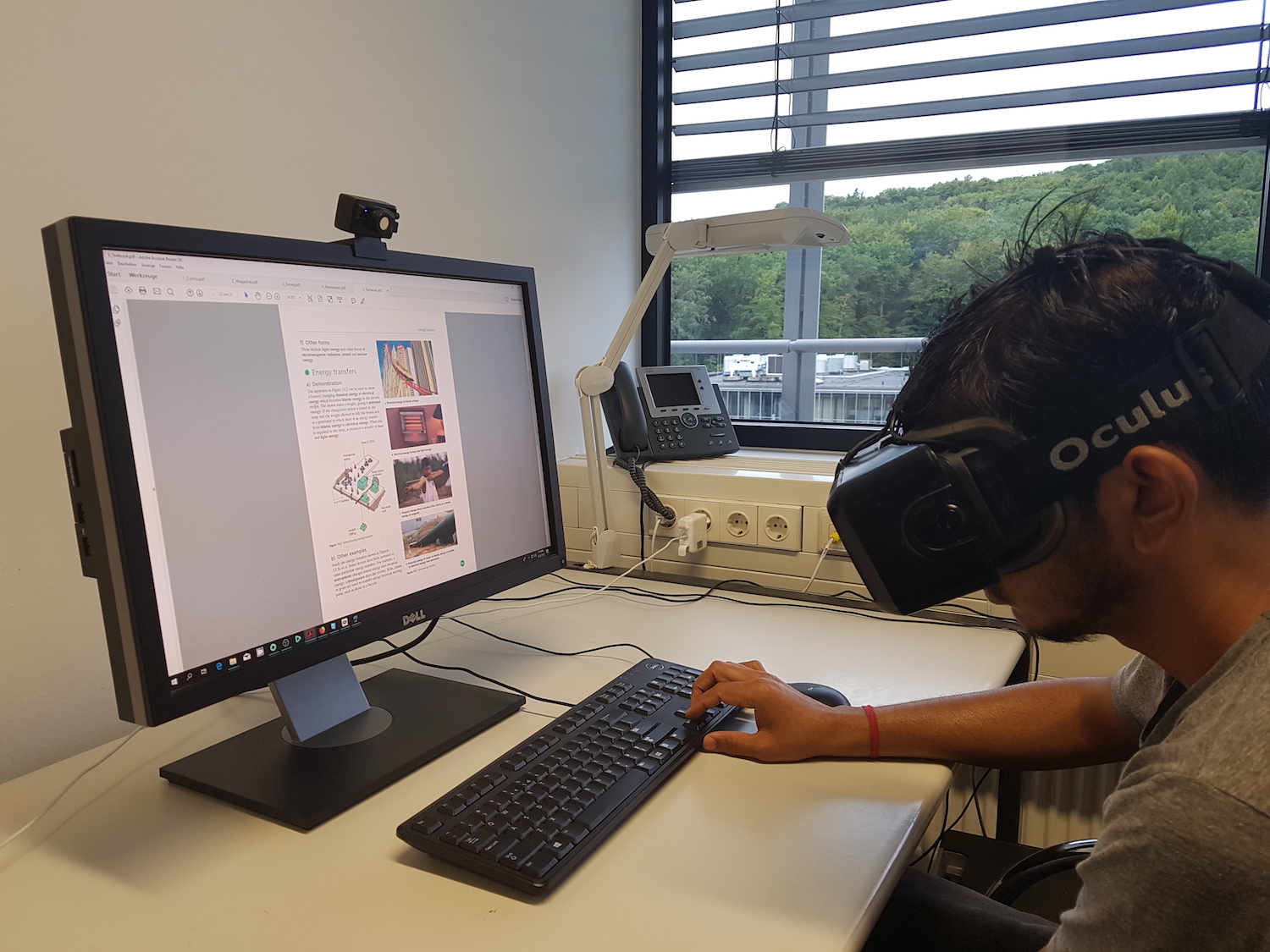}
        \vspace{-0.4cm}
        \caption{Textbook}
        \label{fig:Textbook}
    \end{subfigure}
    \vspace{-0.35cm}
        \caption{Each participant read three different documents: (a) comic, (b) online newspaper, and (c) textbook.}
        \label{fig:DocumentTypes}
    \vspace{-0.25cm}
   \end{figure}

Given the lack of a suitable dataset for evaluating privacy-preserving eye tracking using differential privacy, we recorded our own dataset.
As a utility task, we opted to detect different document types the users 
read, similar to a reading assistant~\cite{kunze2013know}. Instead of printed documents, participants read in VR, wearing a corresponding headset.
The recording of a single participant consists of three separate recording sessions,
in which  
a participant reads one out of three different documents: a comic, online newspaper, or textbook (see Figure~\ref{fig:DocumentTypes}). 
All documents include a varying proportion of text and images. 
Each of these documents  
was about a 10-minute read,
depending on a user's reading skill \mbox{(about 30 minutes in total)}. 

\paragraph{Participants}
We recruited 20 participants (10 male, 10 female) aged 21 to 45 years through university mailing lists and adverts in different university buildings on campus.
Most participants were BSc and MSc students from a large range of subjects (e.g. language science, psychology, business administration, computer science) and different countries (e.g. India, Pakistan, Germany, Italy).
All participants had little or no experience, with eye tracking studies and had normal or corrected-to-normal vision
(contact lenses).

\paragraph{Apparatus}

The recording system consisted of a desktop computer running Windows 10, a 24" computer screen,
and an Oculus DK2 virtual reality headset connected to the computer via USB. 
We fitted the headset with a Pupil eye tracking add-on~\cite{Kassner14_ubicomp} that provides state-of-the-art eye tracking capabilities.
To have more flexibility in the applications used by the participants in the study, we opted for the Oculus
``Virtual Desktop'' that shows arbitrary application windows in the virtual environment.
To record a user's eye movement data, we used the capture software provided by Pupil.
We recorded a separate video from each eye and each document.
Participants used the mouse to start and stop the document interaction 
and were free to read the documents in arbitrary order.
We encouraged participants to read at their usual speed and did not tell them what exactly we were measuring.

\paragraph{Recording Procedure}

After arriving at the lab, participants were 
given time to familiarise themselves with the VR system.
We showed each participant how to behave in the VR environment, given that most of them had never worn a VR headset before.
We did not calibrate the eye tracker but only analysed users' eye movements from the eye videos post hoc.
This was so as not to make participants feel observed, and to be able to record natural 
eye movement 
behaviour.
Before starting the actual recording, we asked participants to sign a consent form.
Participants then started to interact with the VR interface, in which they were asked to read three documents floating in front of them (see Figure~\ref{fig:DocumentTypes}).
After finishing reading a document, the experimental assistant stopped and saved the recording and asked participants questions on their current level of fatigue, whether they liked and understood the document, and whether they found the document difficult
using a 5-point Likert scale (1: Strongly disagree to 5: Strongly agree).
Participants were further asked 
five 
questions about each document to measure 
their
text understanding.
The VR headset was kept on throughout the recording.

After the recording, we asked participants to complete a questionnaire on demographics and any vision impairments.
We also assessed their Big Five personality traits~\cite{john1999big} using established questionnaires from psychology.
In this work we only use the given ground truth information of a user's gender from all collected (private) information, the document type, and IDs we assigned to each participant, respectively.

\paragraph{Eye Movement Feature Extraction}

We extracted a total of 52 eye movement features, covering fixations, saccades, blinks, and pupil diameter (see Table 1 in the supplementary material).
Similar to~\cite{bulling11_pami}, we also computed wordbook features that encode sequences of $n$ saccades.
We extracted these features using a sliding window of 30 seconds (step size of 0.5 seconds).

\section{Evaluation}

The overall goal of our evaluations was to study the effectiveness of the proposed differential privacy method and its potential as a building block for privacy-aware eye tracking.
In these evaluations, gaze-based document type classification served as the utility task, while gender prediction exemplified an attacker without prior knowledge about the target, and user re-identification an attacker with prior knowledge. 

\subsection{Classifier Training}
For each task, we trained a support vector machine (SVM) classifier with radial basis function (RBF) kernel and bias parameter $C=1$ on the extracted eye movement features.
We opted for an SVM due to the good performance demonstrated in a large body of work for eye-based activity recognition ~\cite{bulling11_pami,steil2015discovery}. As the first paper of its kind, one goal was to enable readers to compare our results to the state of the art.
We standardised the training data (zero mean, unit variance) before training the classifiers;
the test data was standardised with the same parameters.
Majority voting was used to summarise all classifications from different time points for the respective participant. 
We randomly sampled training and test sets with an equal distribution of samples for each of the respective classes, 
i.e. for the three document classes, two gender classes and 20 classes for user identification. 

\paragraph{Document Type Classification}

We trained a multi-class SVM for document type classification
and
used leave-one-person-out cross-validation, i.e.\ we trained on the data of 19 participants and tested on the remaining one -- iteratively over all combinations -- and averaged the performance results in the end. 
We envision that in the future, only differentially private data will be available; therefore, we applied our privacy-preserving mechanism to the training and test sets. 
However, currently there is non-noised data available as well: thus, we set up an additional experiment using clean data for training and noised data for testing.

\paragraph{Gender Prediction}
We trained a binary SVM for gender prediction, using reported demographics as ground truth, and applied it again with a person-independent (leave-one-person-out) cross-vali- dation.
Since we are in the \textit{without prior knowledge} threat model, we trained on differentially private and non-noised data to model 
both the future and current situation, as for document type classification.

\paragraph{User (Re-)Identification}
We trained a multi-class SVM for user (re-)identification
but without a leave-one-person-out evaluation scheme.
Instead, we used the first half of the extracted aggregated feature vectors from each document and each participant for training.
We tested on the remaining half, since here we are in the \textit{with prior knowledge} threat model.
In this scenario, we assumed a powerful attacker that was able to obtain training data from multiple people without noise and was able to map 
their samples to their identities. The attacker's goal was to re-identify these people when given noised samples without identity labels.

\paragraph{Implementing the Differential Privacy Mechanism}
We applied the exponential mechanism for each of our $n=20$ participants and for each of the $\numFeatures = 52$ features, 
using a subsampling window size $\subsampleWindow = 10$ to reduce sensitivity. 
In preliminary evaluations, we observed that subsampling alone had no negative effect on the performance of the SVM.
The sensitivity for our differentially private mechanism was
generated by data-driven constraints: For each feature $i$, we estimated $\range{i}$ by calculating the global minimum $min_i$
and maximum $max_i$ over all participants and time points and set $\range{i} = max_i - min_i$. 
This way, the sensitivity ensures privacy protection even of outliers.
The noise we added in our study can be understood as reading-task-specific noise. 
For all $f_i$, we used the same $\epsilon_i$ so that the released data of the whole dataset is \mbox{$\sum_{i=1}^{52} \epsilon_i$-private.}

We repeated our experiments five times each and report averaged results to account for random subsampling and noise generation effects.
As a performance metric, we report $Accuracy=\frac{TP+TN}{TP+FP+TN+FN}$, where TP, FP, TN, and FN represent sample-based true positive, false positive, true negative, and false negative counts.

\subsection{Without Prior Knowledge}
\begin{figure}
 \vspace{-0.3cm}
    \centering
        \centering
        \includegraphics[width=\columnwidth]{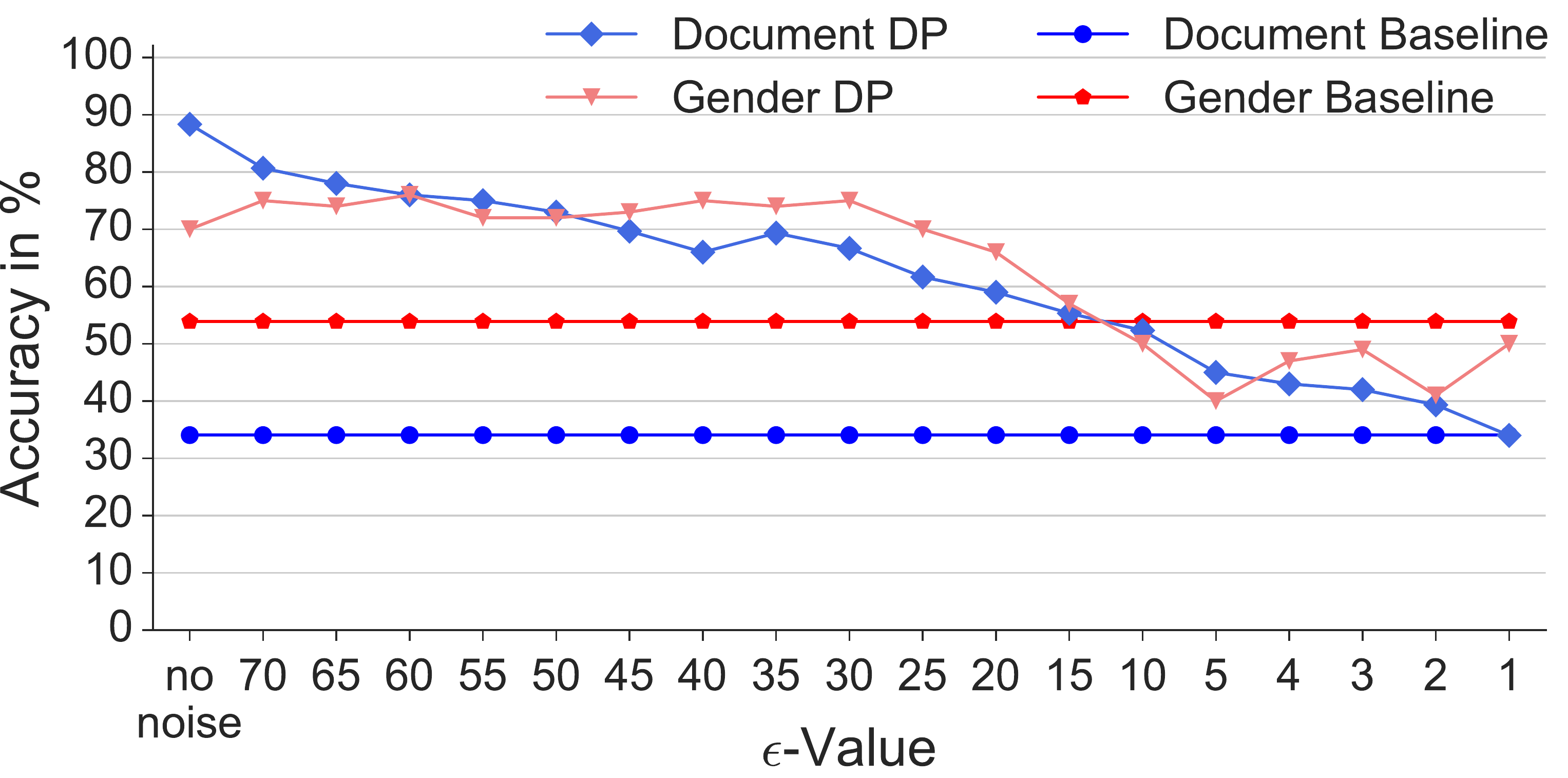} 
        \vspace{-0.75cm}
        \caption{Performance for the threat model without prior knowledge trained on differentially private data.}
        \label{fig:without_prior_knowledge_noised_training}
        \vspace{-0.2cm}
\end{figure}

In Figure~\ref{fig:without_prior_knowledge_noised_training}, we first evaluated the gender prediction task, our example for the attacker \textit{without prior knowledge}, trained on differentially private (noised) data (Gender DP) for decreasing $\epsilon$ values.
As one might expect, decreasing $\epsilon$, and thereby increasing the noise, negatively influences the testing performance 
when trained on differentially private data with $\epsilon < 30$.
For $\epsilon=15$, the performance almost drops to the chance level of 54\% (random guessing in a slightly imbalanced case due to the leave-one-person-out cross-validation).
We conclude that on our dataset, privacy of the participants' gender information is preserved for $\epsilon \leq 15$. 

We then evaluated the impact of the noise level for this $\epsilon$-value on utility (see Figure~\ref{fig:without_prior_knowledge_noised_training})
using the SVMs trained for document type classification on noised data. 
As expected, noise negatively influences document type classification as well, but to a lesser extent compared to gender prediction.
For privacy preservation, it is sufficient to set $\epsilon=15$, resulting in an accuracy of about 55\% for docu- ment type classification, which is still about 22\% over \mbox{chance level.} 

\begin{figure}
  \vspace{-0.3cm}
    \centering
        \centering
        \includegraphics[width=\columnwidth]{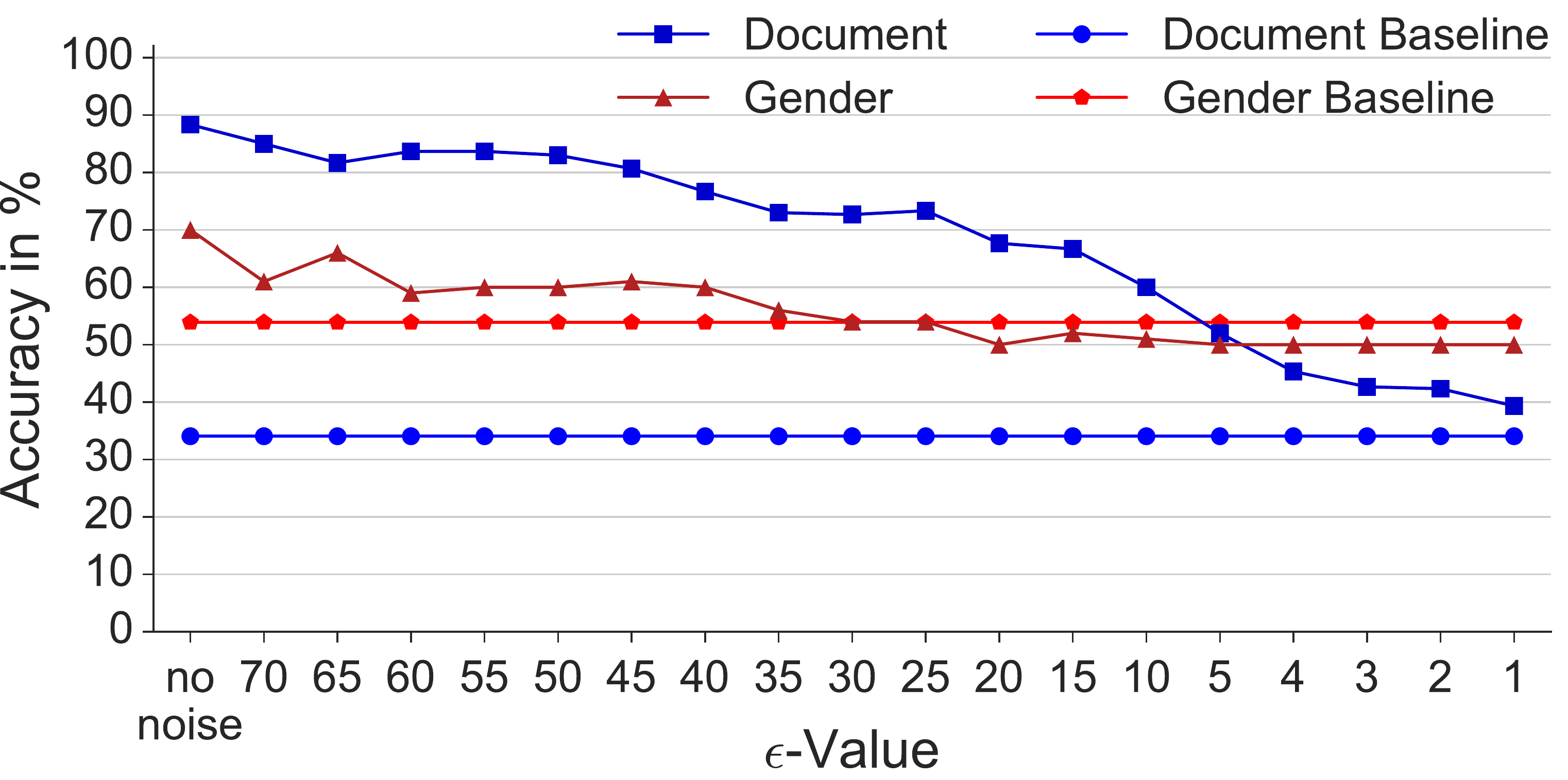} 
        \vspace{-0.75cm}
        \caption{Performance for the threat model without prior knowledge trained on clean data.}
        \label{fig:without_prior_knowledge_clean_training}
        \vspace{-0.2cm}
\end{figure}

So far, we have assumed the SVMs were trained on noised data (Document DP). At present, to the best of our knowledge, all available eye movement datasets are not 
noised.
To study this current situation, we trained both the gender prediction SVM and the document type classification SVM  
without noise and 
tested at various noise levels. Figure~\ref{fig:without_prior_knowledge_clean_training} shows the results of this evaluation.
As can be seen, also in this scenario, privacy can be preserved: 
For $\epsilon=20$, the accuracy of the gender prediction has 
dropped below chance level, while document type classification is still around 70\%. 
We observed that even $\epsilon=30$ would already preserve privacy, since training with noise 
seems to balance out some negative noise effects.
Thus, we conclude that for both current and future situations, privacy preservation is possible while preserving most of the utility.

\begin{figure}
    \centering
        \centering
        \includegraphics[width=\columnwidth]{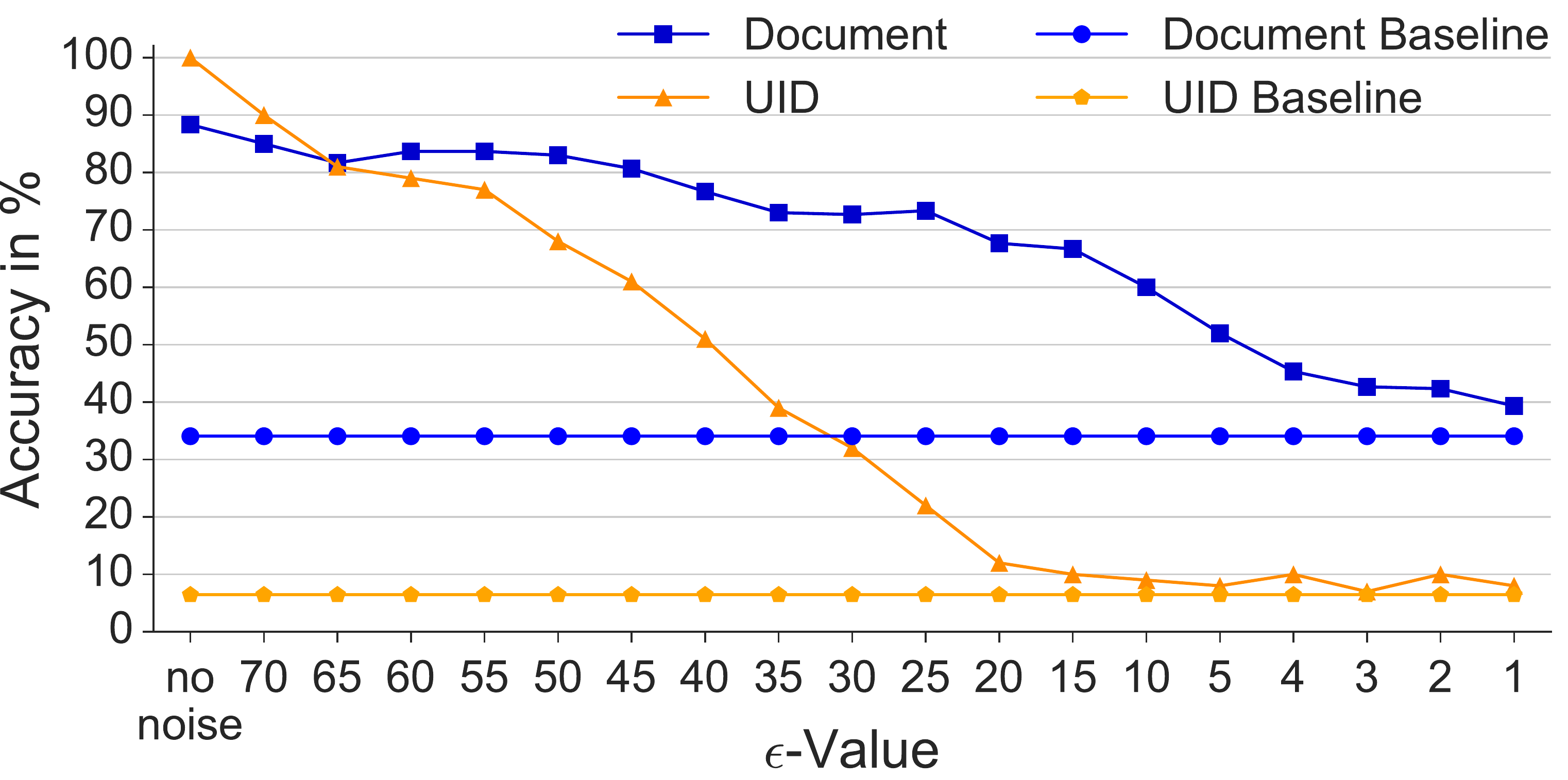} 
        \vspace{-0.75cm}
        \caption{Performance for the threat model with prior knowledge trained on clean data.}
        \label{fig:with_prior_knowledge_clean_training}
        \vspace{-0.2cm}
\end{figure}

\subsection{With Prior Knowledge}

Finally, we evaluated in Figure~\ref{fig:with_prior_knowledge_clean_training} the \textit{with prior knowledge} threat model, in which we assumed the attacker trained a SVM on the data of \mbox{multiple}
users without noise and wanted to re-identify which person a set of noised samples belongs to.
We again added the document type classification performance to be able to judge the effects on utility. 
As expected, the noise on the test data disturbed the attacker's classification ability:
for $\epsilon=40$, the attacker's accuracy dropped to 50\%. For $\epsilon=15$, it dropped down almost to chance level (6.4\%) 
while the utility preserved an accuracy of about 70\%. 
We conclude that, in this scenario as well, it is possible to preserve a user's privacy with acceptable costs on utility.
\vspace{0.1cm}
\section{Discussion}

\subsection{Privacy Concerns in Eye Tracking}

The ever-increasing availability of eye tracking to end users, e.g. in recent VR/AR headsets, in combination with the rich and sensitive information available in the eyes (e.g. on personality~\cite{hoppe2018eye}), creates significant challenges for protecting users' privacy.
Our large-scale online survey on privacy implications of pervasive eye tracking, the first of its kind, yielded a number of interesting insights on this important, yet so far largely unexplored, topic (see the supplementary material for the full results). 
For example, we found that users are willing to share their eye tracking data for medical applications, such as (early) disease detection or stress level monitoring (see Figure~\ref{fig:services}), or for services, if these improve user experience, e.g. in VR or AR (see Figure~\ref{fig:owner}).
On the other hand, participants refused services that use eye movement data for interest identification or shopping assistance, and a majority did not like the idea of services inferring their identity, gender, sexual preference, or race.
These findings are interesting, as they suggest that users are indeed willing to relinquish privacy in return for service use.
They also suggest, however, that users may not be fully aware of the fact that, and to what extent, these services could also infer privacy-sensitive information from their eyes. 
Our proposed differential privacy approach addresses this challenge by allowing sharing of eye movement data while protecting individual privacy.

To prevent inference of users' private attributes from eye tracking data, not every data representation is suitable.
Nonetheless, we identified a clear information gap on the user side, since a majority of participants agreed to share their eye tracking data in almost every data representation (see Figure 3 in the supplementary material).
Participants seemed unaware of the fact that, in particular, raw eye movement data representation is 
inappropriate to protect their privacy. 
Adding noise to this data representation 
would not protect their private attributes either: the added noise could easily be removed by smoothing.
Instead, we recommend using statistical or aggregated feature representations that summarise temporal and appearance statistics of a variety of eye movements, such as fixation, saccades, and blinks.
We are the first to propose a practical solution to this challenge by using differential privacy that effectively protects private information, while at the same time maintaining data utility.

\subsection{Privacy-Preserving Eye Tracking}

Informed by our survey results, we presented a privacy-aware eye tracking method in a VR setting. This is the first of its kind to quantitatively evaluate the practicability and effectiveness of privacy-aware eye tracking. 
For that purpose, we study 1) two realistic threat models (\textit{with} and \textit{without prior knowledge} about the target user), and 2) different scenarios in training with and without clean/non-noised data. We conducted an extensive evaluation on a novel 20-participant dataset and 3) demonstrated the effectiveness of the trained threat models on two example privacy-infringing tasks, namely gender inference and user identification.

Applying differential privacy mitigates these privacy threats.
The fundamental principle of differential privacy is to apply appropriate noise on the data to deteriorate the accuracy of a privacy-infringing task while maintaining that of a utility task. 
As such, the level of noise should be smaller than the inter-class difference in the utility task but larger than that of the privacy-infringing task. 

We showed in our practical evaluations that users' privacy can be preserved with acceptable accuracy of the utility task by applying differential privacy.
This conclusion was consistent across different evaluation paradigms in our example study, which aimed to perform gaze-based document type classification while preserving the privacy of users' gender and identity.

Our mechanism can be used to sanitise data not only before releasing it to the public,
but also in VR/AR devices themselves, since it sanitises one user at a time.
Although our example study focuses only on reading, we expect our method to generalise to any other activity involving eye tracking. Due to our data-driven approach, sensitivity can be adapted so that a similar trade-off can be found. Depending on sensitivity and data vector length, the privacy level $\epsilon$ of this trade-off may differ from the presented  results. Similarly, our study was evaluated on a typical HCI dataset size, and we expect our approach to generalise to larger datasets that will be available in the future, given the rapid emergence of VR and eye tracking technology.

To conclude, the proposed method is an effective and low-cost solution to preserve users' privacy while maintaining the utility task performance.
\section{Conclusion}
In this work we reported the first large-scale online survey to understand users' privacy concerns about eye tracking and eye movement analysis.
Motivated by the findings from this survey, we also 
presented the first privacy-aware gaze interface that uses differential privacy.
We opted for a virtual reality gaze interface, given the significant and imminent threat potential created by upcoming eye tracking technology equipped VR headsets.
Our experimental evaluations on a new 20-participant dataset demonstrated the effectiveness of the proposed approach to preserve private information while maintaining performance on a utility task -- hence, implementing the principle \textit{ensure privacy without impeding utility}.

\begin{acks}
This work was funded, in part, by the \grantsponsor{}{Cluster of Excellence on Multimodal Computing and Interaction (MMCI)}{} at Saarland University, Germany, by a \grantsponsor{}{JST CREST}{} research grant under Grant No.:~\grantnum{1}{JPMJCR14E1}, Japan, as well as by the \grantsponsor{}{German Federal Ministry of Education and Research (BMBF)}{} for the Center for IT-Security, Privacy and Accountability (CISPA) (FKZ:~\grantnum{2}{16KIS0656}).
\end{acks}

\bibliographystyle{ACM-Reference-Format}
\bibliography{references}

\clearpage



\section*{Supplementary material}

## 1 SURVEY RESULTS

We conducted a large-scale online survey to shed light on users' privacy concerns related to eye tracking technology and the information that can be inferred from eye movement data. We advertised our survey on social platforms (Facebook, WeChat) and local mailing lists for study announcements. The survey opened with general questions about eye tracking and VR technologies; continued with questions about future use and applications, data sharing and privacy (especially regarding with whom users are willing to share their data); and concluded with questions about the participants' willingness to share different eye movement representations. Participants answered each question on a 7-point Likert scale (1: Strongly disagree to 7: Strongly agree). To simplify the analysis, we merged scores 1 to 3 to "Disagree" and 5 to 7 to "Agree". At the end we asked for demographic information and offered a raffle.

### 1.1 Services and Attributes

| | Services | | | | | | | | | | | Private Attributes | | | | | |
|---|---|---|---|---|---|---|---|---|---|---|---|---|---|---|---|---|---|
| | Diseases Detection | Natural VR Interaction | Visual Search Target Detection | User Interface Interaction | Understandable Website Content | Reading Skill Improvement | Learning Skill Improvement | Stress Level Monitoring | Interest Identification | Activity Recognition | Shopping Assistance | | Sexual Preference | Gender | Age | Mood and Emotions | Race | Identity |
| 1-3 - Disagree: | 13.71 | 24.19 | 41.94 | 26.61 | 50.81 | 20.16 | 16.13 | 19.35 | 73.39 | 50.81 | 79.03 | | 74.19 | 51.61 | 41.13 | 44.35 | 65.32 | 78.23 |
| 4 - Neither agree nor disagree: | 5.65 | 4.84 | 8.87 | 5.65 | 11.29 | 9.68 | 8.06 | 11.29 | 8.06 | 12.10 | 4.84 | | 6.45 | 7.26 | 12.10 | 12.10 | 8.87 | 4.03 |
| 5-7 - Agree: | 80.65 | 70.97 | 49.19 | 67.74 | 37.90 | 70.16 | 75.81 | 69.35 | 18.55 | 37.10 | 16.13 | | 19.35 | 41.13 | 46.77 | 43.55 | 25.81 | 17.74 |

Fig. 1. Survey results (Services and Attributes): With which services would you agree to share your eye tracking data (Services)?; Would you agree to private attributes being inferred by these services (Private Attributes)?

### 1.2 Whom and Where

| | Sharing | | Owner | | | | | | | | | Environment | | | | Application | |
|---|---|---|---|---|---|---|---|---|---|---|---|---|---|---|---|---|---|
| | Eye Tracking Data | | Governmental Agency (non-health) | Governmental Health Authority | Local Company | International Company | Private Company (user's country) | Private Company (foreign country) | User Himself (home cloud) | Company Internal Use (intranet) | Research Institute | | Public | Private | Constrained | | In Exchange for Benefits | VR/AR |
| 1-3 - Disagree: | 41.13 | | 62.90 | 37.10 | 61.29 | 63.71 | 60.48 | 73.39 | 14.52 | 56.45 | 8.06 | | 63.71 | 58.06 | 32.26 | | 63.71 | 32.26 |
| 4 - Neither agree nor disagree: | 12.90 | | 5.65 | 8.06 | 16.13 | 12.90 | 17.74 | 17.74 | 5.65 | 16.13 | 11.29 | | 9.68 | 16.94 | 13.71 | | 11.29 | 16.13 |
| 5-7 - Agree: | 45.97 | | 31.45 | 54.84 | 22.58 | 23.39 | 21.77 | 8.87 | 79.84 | 27.42 | 80.65 | | 26.61 | 25.00 | 54.03 | | 25.00 | 51.61 |

Fig. 2. Survey results (Whom and Where): Would you agree to share your eye tracking data in general (Sharing); with whom (Owner); where (Environment); in exchange for benefits or for VR/AR usage (Application)?

### 1.3 Data Representation

| | No Modification | | | | | | | | | | | Modified Representation (Anonymised) | | | | | | | | | | |
|---|---|---|---|---|---|---|---|---|---|---|---|---|---|---|---|---|---|---|---|---|---|---|
| | Raw | Temporal Statistics | Appearance Statistics | Fixation Statistics | Fixation Points on Surface | Scan Path on Surface | Saccade Statistics | Gaze Plot | Scan Path Statistics | Heatmaps | Areas of Interest | Aggregated Features | Raw | Temporal Statistics | Appearance Statistics | Fixation Statistics | Fixation Points on Surface | Scan Path on Surface | Saccade Statistics | Gaze Plot | Scan Path Statistics | Heatmaps | Areas of Interest | Aggregated Features |
| 1-3 - Disagree: | 16.94 | 13.71 | 14.52 | 25.00 | 40.32 | 39.52 | 20.97 | 31.45 | 17.74 | 32.26 | 29.84 | 17.74 | 8.87 | 8.87 | 7.26 | 11.29 | 16.94 | 16.13 | 13.71 | 15.32 | 13.71 | 13.71 | 13.71 | 12.90 |
| 4 - Neither agree nor disagree: | 16.13 | 14.52 | 9.68 | 16.13 | 15.32 | 15.32 | 13.71 | 14.52 | 19.35 | 12.90 | 12.10 | 16.13 | 5.65 | 4.84 | 8.06 | 8.06 | 8.87 | 7.26 | 5.65 | 7.26 | 7.26 | 7.26 | 8.87 | 6.45 |
| 5-7 - Agree: | 66.94 | 71.77 | 75.81 | 58.87 | 44.35 | 45.16 | 65.32 | 54.03 | 62.90 | 54.84 | 58.06 | 66.13 | 85.48 | 86.29 | 84.68 | 80.65 | 74.19 | 76.61 | 80.65 | 77.42 | 79.03 | 79.03 | 77.42 | 80.65 |

Fig. 3. Survey results (Data Representations): What kind of data representation would you agree to share (No Modification), and does this behaviour change if the data is anonymised prior to sharing (Anonymised)?

## 2 STATISTICAL TESTS

The following tables correspond to the Figures 1, 2, and 3. We found nearly all answers for the provided questions to be significantly different from equal distribution tested with Pearson's chi-squared test ($p < 0.001$, dof = 6). Additionally, we calculated the skewness and observed that the majority of questions show a significant difference to the corresponding normal distribution ($p < 0.1$).

### 2.1 Statistical tests corresponding to Figure 1

| | Services | | | | | | | | | | | | Private Attributes | | | | | |
|---|---|---|---|---|---|---|---|---|---|---|---|---|---|---|---|---|---|---|
| | Diseases Detection | Natural VR Interaction | Visual Search Target Detection | User Interface Interaction | Understandable Website Content | Reading Skill Improvement | Learning Skill Improvement | Stress Level Monitoring | Interest Identification | Activity Recognition | Shopping Assistance | | Sexual Preference | Gender | Age | Mood and Emotions | Race | Identity |
| chi-squared | 91.19 | 54.84 | 25.48 | 43.1 | 24.92 | 55.63 | 68.84 | 40.73 | 73.58 | 19.5 | 114.9 | | 138.39 | 51.0 | 32.48 | 28.87 | 74.15 | 189.31 |
| p-value | 1.7e-17 | 5.0e-10 | 2.8e-4 | 1.1e-7 | 3.5e-4 | 3.5e-10 | 7.1e-13 | 3.3e-7 | 7.5e-14 | 3.4e-3 | 1.9e-22 | | 2.2e-27 | 3.0e-9 | 1.3e-5 | 6.4e-5 | 5.8e-14 | 3.6e-38 |
| z-score skew | -4.64 | -3.07 | -0.94 | -3.15 | 0.54 | -3.39 | -4.05 | -3.03 | 3.45 | 0.79 | 4.28 | | 4.31 | 0.71 | -1.04 | -0.35 | 2.99 | 4.79 |
| p-value skew | 3.5e-6 | 2.1e-3 | 0.35 | 1.6e-3 | 0.59 | 7.0e-4 | 5.0e-5 | 2.5e-3 | 5.5e-4 | 0.43 | 1.9e-5 | | 1.6e-5 | 0.48 | 0.30 | 0.72 | 2.8e-3 | 1.7e-6 |

### 2.2 Statistical tests corresponding to Figure 2

| | Sharing | | Owner | | | | | | | | | Environment | | | Application | |
|---|---|---|---|---|---|---|---|---|---|---|---|---|---|---|---|---|
| | Eye Tracking Data | | Governmental Agency (non-health) | Governmental Health Authority | Local Company | International Company | Private Company (user's country) | Private Company (foreign country) | User Himself (home cloud) | Company Internal Use (intranet) | Research Institute | | Public | Private | Constrained | In Exchange for Benefits | VR/AR |
| chi-squared | 26.73 | | 44.79 | 22.89 | 43.1 | 55.97 | 61.39 | 77.42 | 72.0 | 26.16 | 88.94 | | 39.03 | 30.0 | 31.81 | 40.5 | 32.26 |
| p-value | 1.6e-4 | | 5.2e-8 | 8.4e-4 | 1.1e-7 | 3.0e-10 | 2.4e-11 | 1.2e-14 | 1.6e-13 | 2.1e-4 | 5.0e-17 | | 7.1e-7 | 4.0e-5 | 1.8e-5 | 3.6e-7 | 1.5e-5 |
| z-score skew | -0.97 | | 1.59 | -1.53 | 1.5 | 1.8 | 1.76 | 2.06 | -4.58 | 1.55 | -4.95 | | 2.1 | 1.7 | -1.97 | 2.55 | -1.99 |
| p-value skew | 0.33 | | 0.11 | 0.13 | 0.13 | 0.07 | 0.08 | 0.04 | 4.7e-6 | 0.12 | 7.5e-7 | | 0.04 | 0.09 | 0.05 | 0.01 | 0.05 |

### 2.3 Statistical tests corresponding to Figure 3

| | No Modification | | | | | | | | | | | | Modified Representation (Anonymised) | | | | | | | | | | |
|---|---|---|---|---|---|---|---|---|---|---|---|---|---|---|---|---|---|---|---|---|---|---|---|
| | Raw | Temporal Statistics | Appearance Statistics | Fixation Statistics | Fixation Points on Surface | Scan Path on Surface | Saccade Statistics | Gaze Plot | Scan Path Statistics | Heatmaps | Areas of Interest | Aggregated Features | Raw | Temporal Statistics | Appearance Statistics | Fixation Statistics | Fixation Points on Surface | Scan Path on Surface | Saccade Statistics | Gaze Plot | Scan Path Statistics | Heatmaps | Areas of Interest | Aggregated Features |
| chi-squared | 41.74 | 57.1 | 63.76 | 29.77 | 13.85 | 9.79 | 40.05 | 21.76 | 48.52 | 19.61 | 26.16 | 49.31 | 95.26 | 96.61 | 91.08 | 74.71 | 51.45 | 58.68 | 75.84 | 61.39 | 68.5 | 68.16 | 62.74 | 73.24 |
| p-value | 2.1e-7 | 1.8e-10 | 7.7e-12 | 4.3e-5 | 3.1e-2 | 1.3e-1 | 4.5e-7 | 1.3e-3 | 9.3e-9 | 3.2e-3 | 2.1e-4 | 6.5e-9 | 2.4e-18 | 1.3e-18 | 1.8e-17 | 4.4e-14 | 2.4e-9 | 8.4e-11 | 2.6e-14 | 2.4e-11 | 8.3e-13 | 9.7e-13 | 1.3e-11 | 8.8e-14 |
| z-score skew | -3.25 | -3.78 | -3.98 | -2.13 | -0.78 | -0.9 | -2.98 | -1.87 | -2.9 | -1.84 | -2.12 | -3.5 | -4.8 | -4.45 | -4.28 | -4.44 | -3.82 | -3.91 | -3.77 | -4.0 | -4.24 | -4.14 | -3.98 | -4.5 |
| p-value skew | 1.1e-3 | 1.6e-4 | 7.0e-5 | 0.03 | 0.44 | 0.37 | 2.9e-3 | 0.06 | 3.7e-3 | 0.07 | 0.03 | 4.7e-4 | 1.6e-6 | 8.6e-6 | 1.9e-5 | 9.1e-6 | 1.3e-4 | 9.4e-5 | 1.6e-4 | 6.4e-5 | 2.2e-5 | 3.4e-5 | 6.9e-5 | 6.6e-6 |

## 3 SURVEY EVALUATION

### 3.1 Eye Tracking and Virtual Reality (VR) Technologies

*3.1.1 I am familiar with eye tracking technology.*

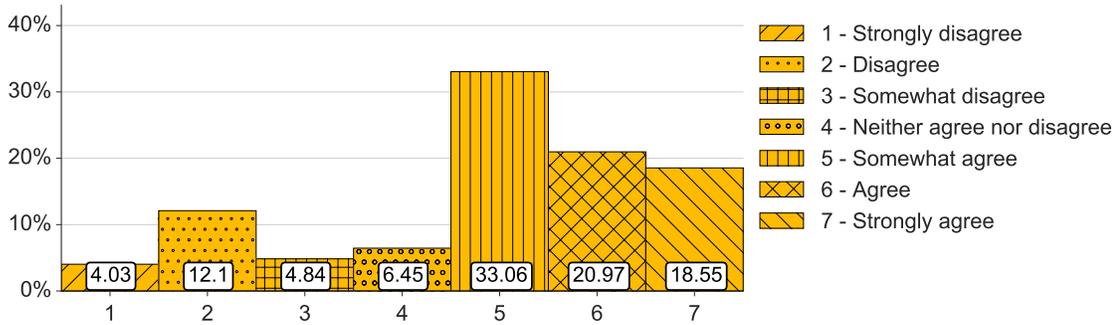

*3.1.2 How many eye tracking applications or experiments have you used or participated?*

*3.1.3 I am concerned about eye tracking technology in terms of ...*

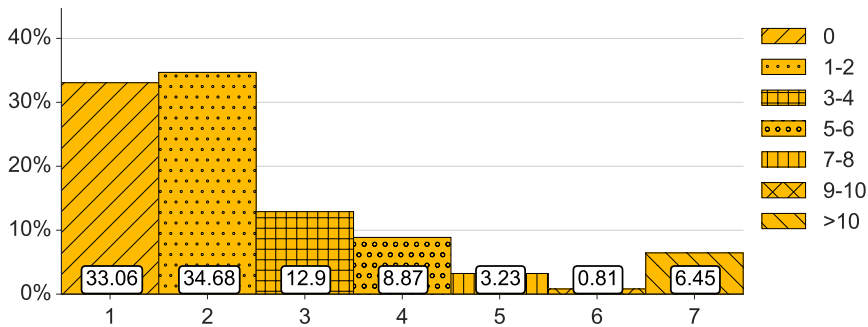

1) social acceptability
   (e.g.: How I am perceived by other people?)
2) mental comfortability
   (e.g.: increase/decrease mental workload)
3) physical comfortability
   (e.g.: increase/decrease physical workload)
4) privacy

*3.1.4   I am familiar with virtual reality (VR) and augmented reality (AR) technology.*

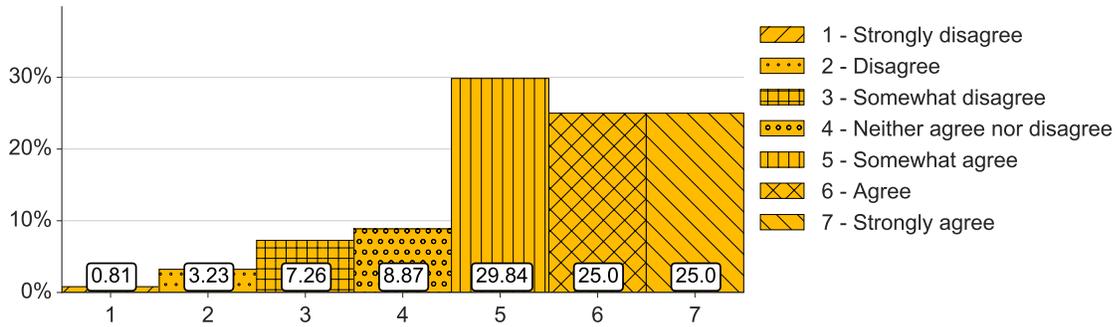

*3.1.5   How many VR applications or experiments have you used or participated?*

*3.1.6   I am concerned about VR technology in terms of ...*

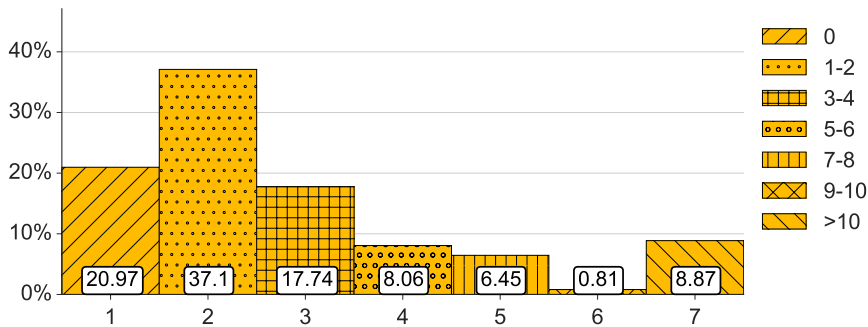

1) social acceptability
   (e.g.: How I am perceived by other people?)
2) mental comfortability
   (e.g.: increase/decrease mental workload)
3) physical comfortability
   (e.g.: increase/decrease physical workload)
4) privacy

### 3.2 Future Use of Eye Tracking Data

*3.2.1 Would you agree to share eye tracking data ...*

| | 1) | 2) | 3) | 4) | 5) | 6) | 7) | 8) | 9) | 10) | 11) |
|---|---|---|---|---|---|---|---|---|---|---|---|
| 1 - Strongly disagree: | 1.61 | 3.23 | 8.87 | 7.26 | 11.29 | 3.23 | 3.23 | 3.23 | 35.48 | 16.13 | 44.35 |
| 2 - Disagree: | 4.84 | 11.29 | 15.32 | 8.06 | 24.19 | 10.48 | 5.65 | 9.68 | 24.19 | 22.58 | 20.97 |
| 3 - Somewhat disagree: | 7.26 | 9.68 | 17.74 | 11.29 | 15.32 | 6.45 | 7.26 | 6.45 | 13.71 | 12.10 | 13.71 |
| 4 - Neither agree nor disagree: | 5.65 | 4.84 | 8.87 | 5.65 | 11.29 | 9.68 | 8.06 | 11.29 | 8.06 | 12.10 | 4.84 |
| 5 - Somewhat agree: | 15.32 | 29.03 | 20.16 | 21.77 | 16.13 | 26.61 | 26.61 | 26.61 | 12.10 | 20.97 | 10.48 |
| 6 - Agree: | 32.26 | 27.42 | 24.19 | 30.65 | 19.35 | 30.65 | 33.06 | 22.58 | 5.65 | 11.29 | 4.84 |
| 7 - Strongly agree: | 33.06 | 14.52 | 4.84 | 15.32 | 2.42 | 12.90 | 16.13 | 20.16 | 0.81 | 4.84 | 0.81 |

1) for early detection of mental and psychological disease like dementia or Parkinson?
2) to enable hands-free interaction with displays and in virtual realty? You could type letters or select a button by gaze interaction or interact with items or persons in VR more naturally.
3) to identify the target of your visual search to provide information about the target you are looking at (e.g. the name of a person, information about a product, etc.)?
4) to improve interactions with user interfaces and devices, e.g. to make them more intuitive or faster?
5) to allow apps and websites to provide content easy to understand?
6) to analyse reading ability and propose methods to improve your reading skills or to change reading material in terms of appearance (enlarging text, highlighting current line)?
7) to analyse and improve your learning skills?
8) monitor your stress level and to provide early-stage healthcare intervention?
9) to identify your interests, e.g. what you like or dislike, and guide you like a shopping assistance, or to steer advertisement?
10) to identify activity specific patterns which could be used for activity tracking, lifelogging or self-quantifying, (e.g. reading, watching TV, playing a video game, computer work, etc.)?
11) to analyse your shopping behaviour on websites or within shopping malls to improve product placement?

*3.2.2 Would you agree to share eye tracking data to identify your ... for a better service (e.g. entertainment, news, business, education, etc.)?*

| | 1) | 2) | 3) | 4) | 5) | 6) |
|---|---|---|---|---|---|---|
| 1 - Strongly disagree: | 49.19 | 31.45 | 20.97 | 25.81 | 37.90 | 55.65 |
| 2 - Disagree: | 18.55 | 15.32 | 12.10 | 10.48 | 20.97 | 17.74 |
| 3 - Somewhat disagree: | 6.45 | 4.84 | 8.06 | 8.06 | 6.45 | 4.84 |
| 4 - Neither agree nor disagree: | 6.45 | 7.26 | 12.10 | 12.10 | 8.87 | 4.03 |
| 5 - Somewhat agree: | 4.84 | 16.13 | 16.94 | 20.16 | 8.06 | 8.06 |
| 6 - Agree: | 12.10 | 20.97 | 26.61 | 18.55 | 14.52 | 9.68 |
| 7 - Strongly agree: | 2.42 | 4.03 | 3.23 | 4.84 | 3.23 | 0.00 |

1) sexual preferences
2) gender
3) age
4) mood and emotions
5) race
6) identity

### 3.3 Sharing Eye Tracking Data

*3.3.1 Would you share your eye tracking data?*

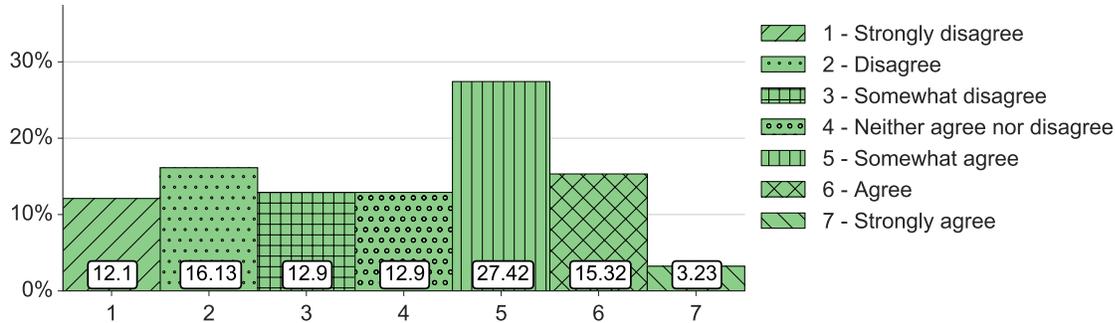

*3.3.2 In general, I would trust a manufacturer and I am willing to share my eye tracking data if it is operated/owned by ...*

| | 1) | 2) | 3) | 4) | 5) | 6) | 7) | 8) | 9) |
|---|---|---|---|---|---|---|---|---|---|
| 1 - Strongly disagree: | 25.00 | 12.10 | 21.77 | 31.45 | 25.81 | 33.06 | 4.03 | 26.61 | 3.23 |
| 2 - Disagree: | 21.77 | 14.52 | 25.00 | 18.55 | 27.42 | 23.39 | 6.45 | 16.94 | 0.81 |
| 3 - Somewhat disagree: | 16.13 | 10.48 | 14.52 | 13.71 | 7.26 | 16.94 | 4.03 | 12.90 | 4.03 |
| 4 - Neither agree nor disagree: | 5.65 | 8.06 | 16.13 | 12.90 | 17.74 | 17.74 | 5.65 | 16.13 | 11.29 |
| 5 - Somewhat agree: | 21.77 | 24.19 | 18.55 | 19.35 | 18.55 | 8.87 | 22.58 | 13.71 | 18.55 |
| 6 - Agree: | 8.87 | 22.58 | 3.23 | 3.23 | 2.42 | 0.00 | 29.84 | 10.48 | 34.68 |
| 7 - Strongly agree: | 0.81 | 8.06 | 0.81 | 0.81 | 0.81 | 0.00 | 27.42 | 3.23 | 27.42 |

1) a governmental agency (non-health related).
2) a governmental health authority (e.g., city, state/province, federal/national).
3) a recognized local company.
4) a recognized international company.
5) a recognized private company in user's country.
6) a recognized private company in foreign country.
7) the user himself (home cloud).
8) company internal use (intranet).
9) research institute.

*3.3.3 Would you share your eye tracking data in exchange for benefits like shopping assistance, activity logging, etc.?*

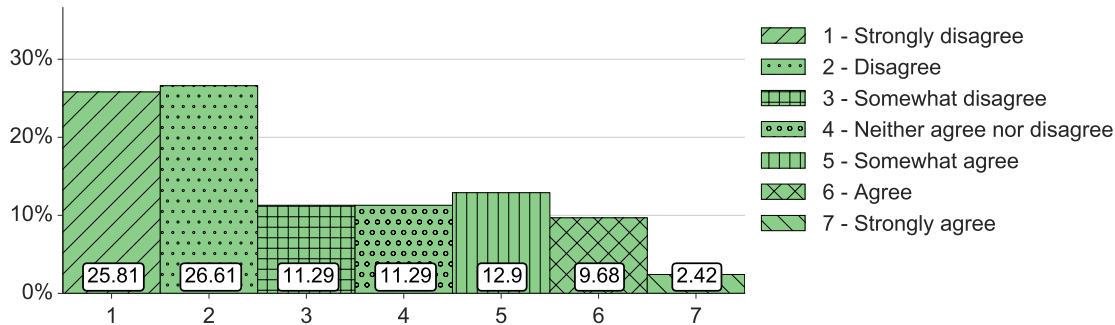

*3.3.4    Would you share your eye tracking data if the data was collected in ...*

| | 1) | 2) | 3) |
|---|---|---|---|
| 1 - Strongly disagree: | 25.81 | 22.58 | 7.26 |
| 2 - Disagree: | 24.19 | 23.39 | 12.10 |
| 3 - Somewhat disagree: | 13.71 | 12.10 | 12.90 |
| 4 - Neither agree nor disagree: | 9.68 | 16.94 | 13.71 |
| 5 - Somewhat agree: | 15.32 | 13.71 | 26.61 |
| 6 - Agree: | 10.48 | 9.68 | 22.58 |
| 7 - Strongly agree: | 0.81 | 1.61 | 4.84 |

1) public, e.g. train station or park?
2) a private environment, e.g. office or home?
3) constrained environments (e.g. a specific room or place)?

*3.3.5    Would you share eye tracking data if the data was collected in one of the following places?*

| | 1) | 2) | 3) | 4) | 5) | 6) | 7) | 8) | 9) | 10) | 11) |
|---|---|---|---|---|---|---|---|---|---|---|---|
| 1 - Strongly disagree: | 29.03 | 25.81 | 25.00 | 23.39 | 16.13 | 60.48 | 29.84 | 12.10 | 32.26 | 33.06 | 34.68 |
| 2 - Disagree: | 25.00 | 25.00 | 22.58 | 20.97 | 20.97 | 19.35 | 26.61 | 15.32 | 19.35 | 20.97 | 20.97 |
| 3 - Somewhat disagree: | 9.68 | 20.97 | 14.52 | 8.06 | 7.26 | 9.68 | 10.48 | 12.10 | 10.48 | 10.48 | 8.87 |
| 4 - Neither agree nor disagree: | 5.65 | 8.06 | 9.68 | 14.52 | 13.71 | 4.84 | 8.06 | 11.29 | 10.48 | 9.68 | 6.45 |
| 5 - Somewhat agree: | 20.16 | 11.29 | 14.52 | 17.74 | 20.16 | 1.61 | 12.90 | 18.55 | 14.52 | 12.90 | 12.10 |
| 6 - Agree: | 8.06 | 8.06 | 12.10 | 10.48 | 15.32 | 3.23 | 10.48 | 19.35 | 11.29 | 10.48 | 14.52 |
| 7 - Strongly agree: | 2.42 | 0.81 | 1.61 | 4.84 | 6.45 | 0.81 | 1.61 | 11.29 | 1.61 | 2.42 | 2.42 |

1) department store
2) friend's home
3) public transport
4) home
5) library
6) restroom
7) workplace
8) car
9) lobby (e.g. hotel)
10) cafe
11) street

*3.3.6    Would you share your eye tracking data if the data was collected in VR or AR?.*

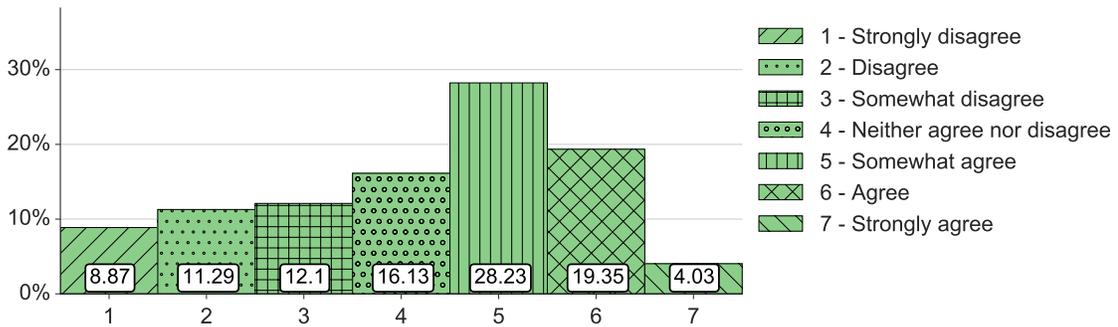

Bar chart values: 8.87, 11.29, 12.1, 16.13, 28.23, 19.35, 4.03

Legend:
1 - Strongly disagree
2 - Disagree
3 - Somewhat disagree
4 - Neither agree nor disagree
5 - Somewhat agree
6 - Agree
7 - Strongly agree

*3.3.7  Would you share your eye tracking data if the data was collected ...*

| | 1) | 2) |
|---|---|---|
| 1 - Strongly disagree: | 12.90 | 13.71 |
| 2 - Disagree: | 20.16 | 17.74 |
| 3 - Somewhat disagree: | 9.68 | 16.13 |
| 4 - Neither agree nor disagree: | 25.00 | 23.39 |
| 5 - Somewhat agree: | 16.13 | 16.13 |
| 6 - Agree: | 12.10 | 11.29 |
| 7 - Strongly agree: | 4.03 | 1.61 |

1) indoor?
2) outdoor?

*3.3.8  Would you share eye tracking data if the collected data was recorded in the ...*

| | 1) | 2) | 3) | 4) | 5) |
|---|---|---|---|---|---|
| 1 - Strongly disagree: | 9.68 | 8.06 | 8.06 | 10.48 | 14.52 |
| 2 - Disagree: | 23.39 | 21.77 | 21.77 | 24.19 | 26.61 |
| 3 - Somewhat disagree: | 7.26 | 5.65 | 4.03 | 6.45 | 11.29 |
| 4 - Neither agree nor disagree: | 31.45 | 30.65 | 33.87 | 34.68 | 29.03 |
| 5 - Somewhat agree: | 12.90 | 17.74 | 16.94 | 10.48 | 12.90 |
| 6 - Agree: | 11.29 | 12.10 | 11.29 | 10.48 | 4.84 |
| 7 - Strongly agree: | 4.03 | 4.03 | 4.03 | 3.23 | 0.81 |

1) morning?
2) noon?
3) afternoon?
4) evening?
5) night?

*3.3.9  Would you share eye tracking data if the recording duration was restricted to one of the following options?*

| | 1) | 2) | 3) | 4) | 5) | 6) | 7) | 8) | 9) |
|---|---|---|---|---|---|---|---|---|---|
| 1 - Strongly disagree: | 1.61 | 10.48 | 14.52 | 22.58 | 8.06 | 17.74 | 18.55 | 20.16 | 44.35 |
| 2 - Disagree: | 7.26 | 17.74 | 20.97 | 30.65 | 8.06 | 23.39 | 23.39 | 29.84 | 25.00 |
| 3 - Somewhat disagree: | 5.65 | 23.39 | 12.90 | 12.90 | 12.10 | 20.16 | 12.90 | 12.10 | 16.13 |
| 4 - Neither agree nor disagree: | 9.68 | 7.26 | 16.94 | 10.48 | 16.13 | 12.10 | 20.16 | 18.55 | 5.65 |
| 5 - Somewhat agree: | 29.03 | 29.03 | 18.55 | 12.90 | 21.77 | 11.29 | 14.52 | 8.87 | 5.65 |
| 6 - Agree: | 33.87 | 10.48 | 10.48 | 8.06 | 20.16 | 11.29 | 7.26 | 4.84 | 2.42 |
| 7 - Strongly agree: | 12.90 | 1.61 | 5.65 | 2.42 | 13.71 | 4.03 | 3.23 | 5.65 | 0.81 |

1) for a specific application with user allowance
2) automatic data recording if eye tracking data is necessary for usage
3) selected hours per day
4) during work time (at work)
5) for personal use
6) during free time
7) during work days
8) during weekend
9) whole day recording

*3.3.10  Would you share eye tracking data if the collected data was saved for ...*

| | 1) | 2) | 3) | 4) | 5) | 6) | 7) | 8) |
|---|---|---|---|---|---|---|---|---|
| 1 - Strongly disagree: | 4.84 | 46.77 | 13.71 | 18.55 | 29.03 | 37.90 | 43.55 | 62.90 |
| 2 - Disagree: | 6.45 | 23.39 | 13.71 | 18.55 | 20.16 | 24.19 | 22.58 | 16.13 |
| 3 - Somewhat disagree: | 10.48 | 8.06 | 13.71 | 12.90 | 13.71 | 8.06 | 12.10 | 8.06 |
| 4 - Neither agree nor disagree: | 12.10 | 9.68 | 20.16 | 20.97 | 12.90 | 12.90 | 9.68 | 7.26 |
| 5 - Somewhat agree: | 35.48 | 8.87 | 20.16 | 14.52 | 15.32 | 10.48 | 7.26 | 5.65 |
| 6 - Agree: | 20.16 | 3.23 | 14.52 | 10.48 | 7.26 | 5.65 | 4.03 | 0.00 |
| 7 - Strongly agree: | 10.48 | 0.00 | 4.03 | 4.03 | 1.61 | 0.81 | 0.81 | 0.00 |

1) application specific purpose (e.g. direct application feedback, VR interaction and gaming, etc)?
2) an unspecified amount of time?
3) a day?
4) a week?
5) a month?
6) a half year?
7) a year?
8) forever?

*3.3.11  Would you share eye tracking data if the data was collected during one of the following emotions?*

| | 1) | 2) | 3) | 4) | 5) | 6) | 7) | 8) |
|---|---|---|---|---|---|---|---|---|
| 1 - Strongly disagree: | 14.52 | 15.32 | 15.32 | 11.29 | 11.29 | 16.13 | 13.71 | 14.52 |
| 2 - Disagree: | 21.77 | 20.97 | 23.39 | 14.52 | 12.90 | 20.16 | 16.94 | 13.71 |
| 3 - Somewhat disagree: | 8.87 | 6.45 | 12.10 | 5.65 | 4.03 | 6.45 | 3.23 | 4.03 |
| 4 - Neither agree nor disagree: | 22.58 | 25.00 | 23.39 | 21.77 | 24.19 | 25.81 | 25.81 | 25.81 |
| 5 - Somewhat agree: | 14.52 | 13.71 | 12.10 | 23.39 | 24.19 | 15.32 | 18.55 | 19.35 |
| 6 - Agree: | 16.13 | 16.94 | 12.10 | 17.74 | 17.74 | 15.32 | 16.94 | 16.94 |
| 7 - Strongly agree: | 1.61 | 1.61 | 1.61 | 5.65 | 5.65 | 0.81 | 4.84 | 5.65 |

1) fear
2) anger
3) sadness
4) joy
5) surprise
6) disgust
7) trust
8) anticipation

*3.3.12	Would you share eye tracking data if it was collected to run an application of the following categories?*

| | 1) | 2) | 3) | 4) | 5) | 6) | 7) | 8) | 9) | 10) | 11) | 12) | 13) | 14) |
|---|---|---|---|---|---|---|---|---|---|---|---|---|---|---|
| 1 - Strongly disagree: | 22.58 | 17.74 | 8.06 | 20.97 | 12.90 | 21.77 | 24.19 | 25.00 | 15.32 | 19.35 | 13.71 | 9.68 | 21.77 | 3.23 |
| 2 - Disagree: | 18.55 | 19.35 | 11.29 | 21.77 | 14.52 | 20.97 | 29.84 | 25.00 | 12.10 | 20.16 | 13.71 | 5.65 | 22.58 | 1.61 |
| 3 - Somewhat disagree: | 17.74 | 20.97 | 11.29 | 20.16 | 11.29 | 14.52 | 18.55 | 15.32 | 8.06 | 14.52 | 12.10 | 6.45 | 11.29 | 3.23 |
| 4 - Neither agree nor disagree: | 13.71 | 12.10 | 10.48 | 13.71 | 12.90 | 13.71 | 8.87 | 15.32 | 12.90 | 15.32 | 17.74 | 8.87 | 17.74 | 5.65 |
| 5 - Somewhat agree: | 20.16 | 16.94 | 25.81 | 16.13 | 25.81 | 17.74 | 10.48 | 11.29 | 32.26 | 16.13 | 17.74 | 22.58 | 12.90 | 15.32 |
| 6 - Agree: | 7.26 | 11.29 | 22.58 | 5.65 | 18.55 | 8.87 | 6.45 | 7.26 | 14.52 | 11.29 | 20.97 | 33.06 | 11.29 | 34.68 |
| 7 - Strongly agree: | 0.00 | 1.61 | 10.48 | 1.61 | 4.03 | 2.42 | 1.61 | 0.81 | 4.84 | 3.23 | 4.03 | 13.71 | 2.42 | 36.29 |

1) utilities (e.g. taxi app, bank app, etc.)
2) entertainment (e.g. streaming, chatting, watching videos, etc.)
3) games (e.g. VR)
4) news
5) productivity
6) lifestyle
7) social networking
8) business
9) education/parenting
10) travel
11) book
12) health/medical and fitness
13) food and drink
14) research (anonymised data storage)

*3.3.13	Would you share eye tracking data if it was collected during one of the following activities?*

| | 1) | 2) | 3) | 4) | 5) | 6) | 7) | 8) | 9) | 10) | 11) | 12) | 13) | 14) |
|---|---|---|---|---|---|---|---|---|---|---|---|---|---|---|
| 1 - Strongly disagree: | 20.97 | 20.16 | 11.29 | 12.90 | 20.16 | 25.81 | 9.68 | 17.74 | 8.87 | 28.23 | 19.35 | 20.97 | 22.58 | 22.58 |
| 2 - Disagree: | 26.61 | 19.35 | 9.68 | 13.71 | 21.77 | 14.52 | 7.26 | 19.35 | 7.26 | 13.71 | 16.94 | 15.32 | 17.74 | 15.32 |
| 3 - Somewhat disagree: | 14.52 | 11.29 | 9.68 | 8.87 | 16.13 | 21.77 | 7.26 | 11.29 | 9.68 | 8.87 | 9.68 | 13.71 | 9.68 | 12.90 |
| 4 - Neither agree nor disagree: | 11.29 | 9.68 | 12.90 | 12.10 | 11.29 | 15.32 | 16.13 | 19.35 | 15.32 | 32.26 | 16.13 | 16.13 | 15.32 | 13.71 |
| 5 - Somewhat agree: | 15.32 | 22.58 | 29.84 | 27.42 | 16.13 | 12.90 | 29.84 | 16.94 | 25.81 | 8.06 | 18.55 | 18.55 | 16.13 | 19.35 |
| 6 - Agree: | 9.68 | 13.71 | 19.35 | 18.55 | 11.29 | 8.06 | 24.19 | 12.10 | 25.00 | 7.26 | 13.71 | 11.29 | 14.52 | 9.68 |
| 7 - Strongly agree: | 1.61 | 3.23 | 7.26 | 6.45 | 3.23 | 1.61 | 5.65 | 3.23 | 8.06 | 1.61 | 5.65 | 4.03 | 4.03 | 6.45 |

1) office work
2) computer work
3) reading
4) writing
5) browsing
6) social interaction
7) gaming
8) eating, drinking
9) driving
10) smoking
11) walking
12) mobile phone interaction
13) watching TV
14) concentrated work

*3.3.14  Would you share eye tracking data if it was collected while you are interacting with one of the following devices?*

|                          | 1) desktop | 2) laptop | 3) mobile | 4) tablet | 5) book | 6) TV |
|---|---|---|---|---|---|---|
| 1 - Strongly disagree:   | 12.90 | 12.90 | 16.13 | 12.90 | 9.68  | 16.94 |
| 2 - Disagree:            | 16.13 | 16.13 | 16.13 | 17.74 | 8.06  | 15.32 |
| 3 - Somewhat disagree:   | 8.87  | 8.87  | 8.87  | 8.06  | 8.06  | 8.06  |
| 4 - Neither agree nor disagree: | 12.90 | 14.52 | 15.32 | 16.94 | 12.10 | 18.55 |
| 5 - Somewhat agree:      | 25.81 | 25.00 | 25.00 | 24.19 | 29.84 | 19.35 |
| 6 - Agree:               | 16.13 | 16.13 | 14.52 | 15.32 | 24.19 | 16.94 |
| 7 - Strongly agree:      | 7.26  | 6.45  | 4.03  | 4.84  | 8.06  | 4.84  |

1) desktop computer  
2) laptop  
3) mobile phone  
4) tablet  
5) book  
6) TV

*3.3.15  Would you share eye tracking data if the data is collected while you are interacting with one of the following persons?*

|                          | 1) friends | 2) relatives | 3) pets | 4) strangers | 5) colleagues | 6) boss |
|---|---|---|---|---|---|---|
| 1 - Strongly disagree:   | 27.42 | 28.23 | 14.52 | 23.39 | 23.39 | 32.26 |
| 2 - Disagree:            | 20.16 | 20.97 | 8.87  | 16.13 | 22.58 | 20.97 |
| 3 - Somewhat disagree:   | 10.48 | 12.10 | 10.48 | 10.48 | 8.06  | 10.48 |
| 4 - Neither agree nor disagree: | 7.26  | 11.29 | 12.10 | 19.35 | 15.32 | 15.32 |
| 5 - Somewhat agree:      | 18.55 | 13.71 | 29.84 | 18.55 | 16.13 | 8.87  |
| 6 - Agree:               | 12.10 | 10.48 | 16.13 | 8.87  | 9.68  | 7.26  |
| 7 - Strongly agree:      | 4.03  | 3.23  | 8.06  | 3.23  | 4.84  | 4.84  |

1) friends  
2) relatives  
3) pets  
4) foreigners/strangers  
5) working colleagues  
6) boss

### 3.3.16 Would you share eye tracking data if the kind of recorded was restricted to ...

| | 1) | 2) | 3) | 4) | 5) | 6) |
|---|---|---|---|---|---|---|
| 1 - Strongly disagree: | 6.45 | 19.35 | 16.13 | 19.35 | 17.74 | 22.58 |
| 2 - Disagree: | 10.48 | 12.90 | 12.90 | 16.13 | 16.13 | 16.13 |
| 3 - Somewhat disagree: | 10.48 | 16.94 | 12.10 | 13.71 | 9.68 | 12.10 |
| 4 - Neither agree nor disagree: | 12.10 | 16.13 | 17.74 | 16.13 | 16.13 | 17.74 |
| 5 - Somewhat agree: | 26.61 | 20.97 | 21.77 | 18.55 | 17.74 | 16.94 |
| 6 - Agree: | 28.23 | 11.29 | 15.32 | 11.29 | 18.55 | 11.29 |
| 7 - Strongly agree: | 5.65 | 2.42 | 4.03 | 4.84 | 4.03 | 3.23 |

1) gaze or pupil behaviour?
2) scene video content?
3) eye video content?
4) gaze or pupil behaviour + scene video content?
5) gaze or pupil behaviour + eye video content?
6) gaze or pupil behaviour + scene video content + eye video content?

### 3.3.17 Suppose you want to create an anonymous online identity in order to share your eye tracking data. Would you "hide" the following ADDITIONAL personal information?

| | 1) | 2) | 3) | 4) | 5) | 6) | 7) | 8) | 9) | 10) | 11) | 12) | 13) | 14) | 15) | 16) | 17) |
|---|---|---|---|---|---|---|---|---|---|---|---|---|---|---|---|---|---|
| 1 - Strongly disagree: | 3.2 | 4.0 | 4.0 | 4.0 | 3.2 | 4.0 | 4.8 | 4.8 | 6.5 | 5.7 | 3.2 | 3.2 | 5.7 | 6.5 | 8.9 | 13.7 | 6.5 |
| 2 - Disagree: | 5.7 | 2.4 | 0.8 | 0.8 | 8.9 | 7.3 | 9.7 | 11.3 | 8.9 | 5.7 | 5.7 | 3.2 | 8.1 | 20.2 | 14.5 | 20.2 | 13.7 |
| 3 - Somewhat disagree: | 5.7 | 2.4 | 2.4 | 2.4 | 6.5 | 8.1 | 16.9 | 19.4 | 16.1 | 14.5 | 4.0 | 3.2 | 13.7 | 14.5 | 12.9 | 19.4 | 15.3 |
| 4 - Neither agree nor disagree: | 5.7 | 3.2 | 3.2 | 2.4 | 10.5 | 15.3 | 18.6 | 16.9 | 12.9 | 13.7 | 8.9 | 2.4 | 13.7 | 15.3 | 16.1 | 14.5 | 12.9 |
| 5 - Somewhat agree: | 8.1 | 9.7 | 6.5 | 3.2 | 10.5 | 12.1 | 13.7 | 16.1 | 10.5 | 8.9 | 6.5 | 4.0 | 10.5 | 9.7 | 7.3 | 8.1 | 6.5 |
| 6 - Agree: | 27.4 | 24.2 | 27.4 | 26.6 | 22.6 | 18.6 | 12.1 | 9.7 | 18.6 | 18.6 | 20.2 | 23.4 | 20.2 | 14.5 | 12.9 | 6.5 | 12.9 |
| 7 - Strongly agree: | 44.4 | 54.0 | 55.6 | 60.5 | 37.9 | 34.7 | 24.2 | 21.8 | 26.6 | 33.1 | 51.6 | 60.5 | 28.2 | 19.4 | 27.4 | 17.7 | 32.3 |

1) first name
2) last name
3) identifiable profile picture
4) residential address
5) city where I live
6) occupation and employment information
7) hobbies
8) interests
9) current location information (e.g., kitchen, public transport, office)
10) my health condition(s)
11) email address
12) phone number
13) age and date of birth
14) gender
15) race
16) eye color
17) iris image

### 3.4 Eye Tracking Data Representations

*3.4.1 Would you agree to share eye tracking data which consists of ...*

| | 1) | 2) | 3) | 4) | 5) | 6) | 7) | 8) | 9) | 10) | 11) | 12) |
|---|---|---|---|---|---|---|---|---|---|---|---|---|
| 1 - Strongly disagree: | 4.03 | 3.23 | 3.23 | 3.23 | 11.29 | 15.32 | 4.84 | 9.68 | 5.65 | 8.87 | 8.06 | 5.65 |
| 2 - Disagree: | 7.26 | 7.26 | 8.06 | 12.10 | 14.52 | 10.48 | 9.68 | 13.71 | 8.06 | 12.10 | 13.71 | 8.87 |
| 3 - Somewhat disagree: | 5.65 | 3.23 | 3.23 | 9.68 | 14.52 | 13.71 | 6.45 | 8.06 | 4.03 | 11.29 | 8.06 | 3.23 |
| 4 - Neither agree nor disagree: | 16.13 | 14.52 | 9.68 | 16.13 | 15.32 | 15.32 | 13.71 | 14.52 | 19.35 | 12.90 | 12.10 | 16.13 |
| 5 - Somewhat agree: | 22.58 | 25.00 | 29.03 | 20.16 | 20.16 | 17.74 | 27.42 | 22.58 | 31.45 | 22.58 | 25.00 | 22.58 |
| 6 - Agree: | 27.42 | 29.84 | 28.23 | 26.61 | 19.35 | 20.16 | 25.00 | 23.39 | 18.55 | 23.39 | 23.39 | 30.65 |
| 7 - Strongly agree: | 16.94 | 16.94 | 18.55 | 12.10 | 4.84 | 7.26 | 12.90 | 8.06 | 12.90 | 8.87 | 9.68 | 12.90 |

1) the raw x and y gaze or pupil position over time?
2) statistics of steady (fixations) and dynamic (saccades) state of the eyes (when fixations and saccades take place)?
3) statistics of steady (fixations) and dynamic (saccades) state of the eyes (how often fixations and saccades appear in a given time range)?
4) statistics of eye tracking data which describe the number of fixations, fixation duration as well as their spatial distribution on public displays, computer monitors, or in VR environment?
5) fixation points on public displays, computer monitors, or in VR environment?
6) scan path information, the concatenation of gaze movements on public displays, computer monitors, or in VR environment?
7) scan path statistics, e.g. whether after a gaze movement to left is followed by a movement upwards, on public displays, computer monitors, or in VR environment?
8) fixations with duration and scan path information (Gaze Plot), the concatenation of gaze movements and fixational behaviour, on public displays, computer monitors, or in VR environment?
9) fixations with duration and scan path information statistics, e.g. whether after a gaze movement to left is followed by a movement upwards and how long the following fixation lasts, on public displays, computer monitors, or in VR environment?
10) heatmaps, user's gaze distribution on public displays, computer monitors, or in VR environment?
11) statistics of gaze distribution on areas of interests (AOIs) on public displays, computer monitors, or in VR environment?
12) aggregated features, given as so-called feature vectors, where each entry of such a vector describe a feature of user's behaviour like average blinking rate, fixation duration, ratio of saccadic movements, etc. within a given time window?

*3.4.2  Imagine your eye tracking data could be modified so that it is anonymous, i.e. indistinguishable from that of another user. Would you agree to share ...*

| | 1) | 2) | 3) | 4) | 5) | 6) | 7) | 8) | 9) | 10) | 11) | 12) |
|---|---|---|---|---|---|---|---|---|---|---|---|---|
| 1 - Strongly disagree: | 1.61 | 0.81 | 0.81 | 2.42 | 4.03 | 3.23 | 0.81 | 3.23 | 2.42 | 3.23 | 2.42 | 3.23 |
| 2 - Disagree: | 2.42 | 3.23 | 3.23 | 4.03 | 5.65 | 7.26 | 6.45 | 5.65 | 5.65 | 4.84 | 6.45 | 4.84 |
| 3 - Somewhat disagree: | 4.84 | 4.84 | 3.23 | 4.84 | 7.26 | 5.65 | 6.45 | 6.45 | 5.65 | 5.65 | 4.84 | 4.84 |
| 4 - Neither agree nor disagree: | 5.65 | 4.84 | 8.06 | 8.06 | 8.87 | 7.26 | 5.65 | 7.26 | 7.26 | 7.26 | 8.87 | 6.45 |
| 5 - Somewhat agree: | 25.00 | 28.23 | 27.42 | 24.19 | 22.58 | 25.81 | 29.03 | 25.81 | 23.39 | 27.42 | 25.00 | 25.00 |
| 6 - Agree: | 32.26 | 29.84 | 29.03 | 28.23 | 25.81 | 25.00 | 23.39 | 25.00 | 28.23 | 23.39 | 25.00 | 27.42 |
| 7 - Strongly agree: | 28.23 | 28.23 | 28.23 | 28.23 | 25.81 | 25.81 | 28.23 | 26.61 | 27.42 | 28.23 | 27.42 | 28.23 |

1) the raw x and y gaze or pupil position over time?
2) statistics of steady (fixations) and dynamic (saccades) state of the eyes (when fixations and saccades take place)?
3) statistics of steady (fixations) and dynamic (saccades) state of the eyes (how often fixations and saccades appear in a given time range)?
4) statistics of eye tracking data which describe the number of fixations, fixation duration as well as their spatial distribution on public displays, computer monitors, or in VR environment?
5) fixation points on public displays, computer monitors, or in VR environment?
6) scan path information, the concatenation of gaze movements on public displays, computer monitors, or in VR environment?
7) scan path statistics, e.g. whether after a gaze movement to left is followed by a movement upwards, on public displays, computer monitors, or in VR environment?
8) fixations with duration and scan path information (Gaze Plot), the concatenation of gaze movements and fixational behaviour, on public displays, computer monitors, or in VR environment?
9) fixations with duration and scan path information statistics, e.g. whether after a gaze movement to left is followed by a movement upwards and how long the following fixation lasts, on public displays, computer monitors, or in VR environment?
10) heatmaps, user's gaze distribution on public displays, computer monitors, or in VR environment?
11) statistics of gaze distribution on areas of interests (AOIs) on public displays, computer monitors, or in VR environment?
12) aggregated features, given as so-called feature vectors, where each entry of such a vector describe a feature of user's behaviour like average blinking rate, fixation duration, ratio of saccadic movements, etc. within a given time window?

*3.4.3 Would you agree to share scene video information of public displays, computer monitors, or in VR environment which consists of ...*

| | 1) | 2) | 3) | 4) | 5) | 6) |
|---|---|---|---|---|---|---|
| 1 - Strongly disagree: | 10.48 | 8.87 | 8.06 | 4.84 | 4.84 | 8.06 |
| 2 - Disagree: | 16.94 | 15.32 | 14.52 | 16.13 | 16.13 | 16.13 |
| 3 - Somewhat disagree: | 21.77 | 13.71 | 14.52 | 13.71 | 12.90 | 11.29 |
| 4 - Neither agree nor disagree: | 16.94 | 20.16 | 16.94 | 17.74 | 17.74 | 14.52 |
| 5 - Somewhat agree: | 19.35 | 23.39 | 25.81 | 25.81 | 25.81 | 28.23 |
| 6 - Agree: | 12.10 | 15.32 | 16.94 | 16.94 | 19.35 | 17.74 |
| 7 - Strongly agree: | 2.42 | 3.23 | 3.23 | 4.84 | 3.23 | 4.03 |

1) the scene content during a whole eye tracking recording?
2) single frame from each fixation?
3) gaze stripes, sequence of image frame from each fixation?
4) tiny image patches around the gaze position during a whole eye tracking recording?
5) tiny image patches around the gaze position from each fixation?
6) the whole scene video but with blurred out surrounding and clear object of interest?

*3.4.4 Would you agree to share eye video information recorded from eye tracking camera which consists of ...*

| | 1) | 2) | 3) |
|---|---|---|---|
| 1 - Strongly disagree: | 17.74 | 14.52 | 6.45 |
| 2 - Disagree: | 20.16 | 18.55 | 15.32 |
| 3 - Somewhat disagree: | 19.35 | 11.29 | 8.06 |
| 4 - Neither agree nor disagree: | 11.29 | 10.48 | 10.48 |
| 5 - Somewhat agree: | 13.71 | 20.97 | 22.58 |
| 6 - Agree: | 13.71 | 16.13 | 27.42 |
| 7 - Strongly agree: | 4.03 | 8.06 | 9.68 |

1) the whole eye video with visible iris and surrounding facial expressions?
2) share the whole eye video with visible iris but blurred surrounding facial expressions?
3) the whole eye video but only showing the pupil centre without iris or facial expressions?

## 4 EYE MOVEMENT FEATURE EXTRACTION

Table 1 summarises the features that we extracted from fixations, saccades, blinks, pupil diameter, and a user's scan paths. Similar to [Bulling et al. 2011], each saccade is encoded as a character forming words of length $n$ (wordbook). We extracted these features on a sliding window of 30 seconds (step size of 0.5 seconds).

| | |
|---|---|
| Fixation (8) | rate, mean, max, var of durations, mean/var of var pupil position within one fixation |
| Saccades (12) | rate/ratio of (small/large/right/left) saccades, mean, max, variance of amplitudes |
| Combined (1) | ratio saccades to fixations |
| Wordbooks (24) | number of non-zero entries, max and min entries, and their difference for n-grams with n <= 4 |
| Blinks (3) | rate, mean/var blink duration |
| Pupil Diameter (4) | mean/var of mean/var during fixations |

Table 1. We extracted 52 eye movement features to describe a user's eye movement behaviour. The number of features per category is given in parentheses.



\end{document}